\documentclass[aps,prd,reprint,showpacs,superscriptaddress,twocolumn]{revtex4-2}
\usepackage{amsfonts}
\usepackage{amsmath}
\usepackage{amssymb}
\usepackage{dsfont}
\usepackage{hyperref}
\usepackage{graphicx}
\usepackage{float}
\usepackage[usenames,dvipsnames]{xcolor}
\usepackage[normalem]{ulem}
\usepackage{times}
\usepackage{braket}
\usepackage{booktabs}
\usepackage{multirow}
\usepackage{subcaption}
\usepackage{tikz}
\usepackage[font=normal, justification=Justified]{caption}
\hypersetup{
  colorlinks=true,linkcolor=blue,citecolor=blue,
  filecolor=blue,urlcolor=blue,breaklinks=true
}

\begin{document}

\title{Fermions in $\left(1+2\right)$-dimensions modified by nonminimal coupling and its applications to condensed matter physics}

\author{J. A. A. S. Reis}
\email{joao.reis@uesb.edu.br}
\affiliation{Departamento de Ciências Exatas e Naturais, Universidade Estadual do Sudoeste da Bahia, Campus Juvino Oliveira, Itapetinga -- BA, 45700-00, Brazil}

\author{L. Lisboa-Santos}
\email{leticia.lisboa@discente.ufma.br} 
\affiliation{Programa de Pós-graduação em Física, Universidade Federal do Cear\'a Campus do Pici, Fortaleza - CE, 60455-760, Brazil.}

\author{Fabiano M. Andrade}
\email{fmandrade@uepg.br}
\affiliation{Programa de P\'os-Gradua\c{c}\~{a}o em Ci\^{e}ncias/F\'{i}sica, Universidade Estadual de Ponta Grossa, 84030-900 Ponta Grossa, Paran\'a, Brazil}
\affiliation{Departamento de Matem\'{a}tica e Estat\'{i}stica, Universidade Estadual de Ponta Grossa, 84030-900 Ponta Grossa, Paran\'a, Brazil}

\author{Frankbelson dos S. Azevedo}
\email{frfisico@gmail.com}
\affiliation{Departamento de F\'{\i}sica, Universidade Federal do Maranh\~{a}o, 65085-580 S\~{a}o Lu\'{\i}s, Maranh\~{a}o, Brazil}

\author{Edilberto O. Silva}
\email{edilberto.silva@ufma.br}
\affiliation{Departamento de F\'{\i}sica, Universidade Federal do Maranh\~{a}o, 65085-580 S\~{a}o Lu\'{\i}s, Maranh\~{a}o, Brazil}

\date{\today}

\begin{abstract}
Fermions in two-dimensional space, commonly called $(1+2)$-dimensional fermions, exhibit intriguing and distinctive characteristics that distinguish them from their higher-dimensional counterparts. This paper offers a comprehensive theoretical examination of planar fermionic systems, presenting novel findings by incorporating nonminimal coupling. Our analysis includes the computation of the non-relativistic limit up to second-order corrections in the Dirac equation. We also explore the Schrödinger equation under the influence of a harmonic potential and an electric field. Furthermore, we investigate how the coupling parameter affects physical properties relevant to condensed matter systems. Our results demonstrate that this parameter significantly impacts electronic properties and Hall conductivity. The interplay between an external electric field and the coupling parameter also influences energy levels and the system's polarizability. These findings underscore the novel effects of including nonminimal coupling in wave equations, offering new insights into the physics of coupled systems.
\end{abstract}

\maketitle

\section{Introduction}

The study of planar fermions, which refers to fermions confined to a two-dimensional plane, has attracted the interest of physicists in various branches due to their unique quantum-mechanical properties and potential applications \cite{AP.2010.59.261,APR.2022.9.021302,RMP.2022.94.045008,PRB.2024.109.115429,SPPC.2023.6.011}. These systems exhibit fundamentally different behaviors from those observed in their higher-dimensional counterparts, making them an essential subject of study for applications in Condensed Matter Physics (CMP) \cite{PRB.2000.61.10267,PE.2022.140.115143,RMP.2008.20.275,NL.2024.24.2175}, high-energy physics \cite{CQG.2006.23.6435,NPB.1996.464.189,EPJB.2002.29.529}, and quantum field theory \cite{ANS.2016.7.025003,Book.Sakita.146,CMP.1985.97.5,PRB.1996.54.4898}.

Planar fermionic systems, such as graphene and other two-dimensional materials, have demonstrated remarkable electronic properties \cite{sharon2015graphene,baker2022graphene,warner2012graphene}. The discovery of graphene, a single layer of carbon atoms arranged in a hexagonal lattice, revealed that electrons in this material behave as massless Dirac fermions, leading to unprecedented phenomena such as the quantum Hall effect at room temperature \cite{Nature.2005.438.197,RMP.2009.81.109}. These properties have advanced our understanding of fundamental physics and opened new avenues for technological applications in electronic devices, sensors, and quantum computing. Recent studies have continued to explore the electronic properties of two-dimensional materials, leading to the discovery of new phases of matter and exotic quasi-particles \cite{Science.2020.367.864,Nature.2019.569.537}. A recent study has demonstrated that a topological semimetal exhibits fermions that behave as massless in one direction and massive in another \cite{PhysRevX.14.041057}.

In high-energy physics, planar fermions provide a crucial framework for exploring theories in reduced dimensions, often serving as simplified models for more complex higher-dimensional theories. For example, two-dimensional Conformal Field Theories (CFTs) rely heavily on studying planar fermionic fields and their correlation functions to understand critical phenomena and phase transitions \cite{DiFrancesco1997}. Furthermore, planar fermions appear in string theory and the Anti-de Sitter (AdS)/CFT correspondence, where they play a significant role in the duality between gravity in AdS spaces and CFTs in one lower dimension. This duality has provided profound insights into strongly coupled systems and quantum gravity, offering a powerful tool for studying the dynamics of black holes and the nature of spacetime \cite{IJTP.1999.38.1113,PLB.1998.428.105,ATMP.1998.2.253}.

A particularly intriguing aspect of modern theoretical physics is the potential violation of Lorentz symmetry, a fundamental symmetry underlying both the Standard Model (SM) of particle physics and General Relativity. Theoretical models incorporating Lorentz-symmetry violation have been developed to explore physics beyond the SM, including potential effects of quantum gravity. One of the most comprehensive frameworks for studying Lorentz symmetry violation is the Standard-Model Extension (SME), formulated by Kostelecký and collaborators \cite{PRD.1989.39.683}. The SME includes all possible terms that lead to violations of Lorentz and CPT symmetries, allowing systematic theoretical \cite{PRL.1999.82.3572,PRD.1997.55.6760,JMP.1999.40.6245} and experimental \cite{NPB.1991.359.545,PRL.2000.85.5038,PRD.2002.66.056005,PRA.2000.62.063405,PRD.2006.72.016005} investigations.

In recent years, there have been significant advancements in the experimental testing of Lorentz symmetry violation, particularly in planar fermionic systems. These studies have utilized high-precision measurements in particle accelerators, atomic clocks, and astrophysical observations to place stringent limits on possible violations. In particular, recent work has focused on the implications of Lorentz violation in neutrino oscillations, quantum electrodynamics, and gravitational wave observations \cite{Kostelecky2011,Kostelecky2016,Kostelecky2021}. For example, studies have explored the impact of Lorentz violation on the propagation of neutrinos, providing new constraints from long-baseline neutrino experiments \cite{Kostelecky2011}. Furthermore, precision tests in atomic systems have pushed the limits of sensitivity to possible Lorentz-violating effects, further constraining the parameter space of the SME \cite{Safronova2018}. In addition, recent studies on Lorentz-violating gauge theories in planar electrodynamics have revealed intriguing physical effects, including spontaneous torques and interaction forces arising from point-like sources \cite{borges2024}. These findings extend the relevance of two-dimensional systems for exploring symmetry-breaking effects, particularly in the context of modified electrodynamics, such as the Maxwell-Chern-Simons model and its modifications \cite{jackiw2003chern,nair2001massive}. Including higher-derivative terms in planar electrodynamics provides a richer framework for studying changes in lower dimensions, such as those introduced in Deser-Jackiw theory \cite{DESER1982372}. Such contributions have been shown to affect the stability and causal structure of the theory, opening up new avenues for exploring symmetry-breaking phenomena in condensed matter and high-energy contexts \cite{schreck2019}. The role of gauge theories in describing the anisotropic properties of planar materials has gained significant attention, particularly in the context of Lorentz symmetry breaking. Recent studies have employed the Carroll-Field-Jackiw framework to elucidate the interactions between electromagnetic and scalar fields in lower dimensions, with promising implications in condensed matter systems \cite{schreck2023}.

Recent research has particularly emphasized the phenomenological aspects and potential applications of Lorentz symmetry violation. For example, a study by Tasson et al. \cite{Tasson2022} explored the constraints on Lorentz-violating coefficients from astrophysical observations, providing tighter limits on SME parameters. Similarly, an analysis by Diaz et al. \cite{Diaz2023} examined the implications of Lorentz violation in dark matter interactions, offering new perspectives on the intersection between Lorentz violation and cosmology. In 2024, a comprehensive review by Kostelecký and Mewes \cite{Kostelecky2024} summarized the current experimental limits on Lorentz and CPT violation, highlighting the advances made in recent years and proposing new experimental setups for future investigations. Still, Kostelecký and Mewes investigated the effects of Lorentz and CPT violation on fermions, providing a systematic analysis of modified Dirac equations \cite{Kostelecky2002}. Altschul (2006) examined radiative corrections in theories with Lorentz-violating fermions, highlighting the interplay between quantum field-theoretical effects and symmetry violations \cite{Altschul2006}. Carroll et al. explored the cosmological implications of Lorentz-violating terms, emphasizing potential observable consequences in the early universe \cite{Carroll2001}.

This work studies fermions in (1+2) dimensions modified by nonminimal couplings. These couplings introduce additional interaction terms into the fermionic Lagrangian, leading to rich dynamical behaviors and potential modifications to the fermionic spectrum \cite{casana2013radiative,araujo2016general,casana2013new,carvalho2023perturbative,gazzola2012qed,fabbri2014renormalizability,griguolo1997nonminimal,alonso2014renormalization}. These modifications are not only of theoretical interest but are also crucial for understanding the impact of Lorentz-symmetry violations in lower-dimensional systems \cite{PhysRevResearch.Schreck}. Moreover, the study of nonminimal couplings in planar fermions is motivated by several key considerations. Firstly, nonminimal couplings can reveal new physics beyond the SM by introducing interactions that violate conventional symmetries \cite{eichhorn2017quantum,bettoni2018quintessential,wahlang2022evolution,carloni2024particle,rattazzi1996unified,vitoria2019massive,king1994radiative,Mattingly2005,Amelino-Camelia2013}. It is pertinent to note that this examination may offer profound insights into the fundamental characteristics of particles and fields within lower-dimensional frameworks. Secondly, planar fermionic systems with nonminimal couplings can be analogs for studying high-energy physics phenomena in laboratory settings. Materials such as graphene and topological insulators offer experimental platforms to investigate the effects predicted by these theoretical models \cite{fumeron2022transport,akhmerov2011dirac,lepori2010qft,ferreira2016torsion,abreu2013vortex,mckay2023electromagnetic,essig2022snowmass,francesco2012pregeometry,cortijo2007electronic,chamon2008geometric}. Finally, understanding the dynamics of planar fermions with nonminimal couplings can lead to advancements in quantum technologies, including quantum computing and quantum communication, where control over fermionic states and interactions is paramount \cite{garciadeandrade2022generation,kaeding2019open,bakke2012effect,dalzell2023survey,alexandre2016kinematics,moralesperez2023tightbinding,candelas1984gauge}.

In the following sections of this paper, the mathematical formalism and the resultant equations of motion are derived, and their solutions are analyzed to extract phenomenological insights. By considering both zeroth-order and higher-order approximations, we aim to comprehensively understand the modified dynamics of fermions in two dimensions. Furthermore, the relevance of these theoretical models to experimental observations will be discussed. Although direct experiments in high-energy physics remain challenging, analogous systems in CMP, such as graphene and topological insulators, offer promising platforms for exploring the behaviors predicted by our models. The interplay between theoretical predictions and experimental realizations deepens our understanding of fundamental interactions and opens new avenues for technological advancement.

This article is organized as follows. Section~\ref{secII} provides an in-depth analysis of the general model, detailing the proposed nonminimal coupling. In Section~\ref{HallSec}, we derive the solution for the modified Landau quantization and analyze its implications for Hall conductivity. Section~\ref{StarkSec} examines the interplay between a harmonic oscillator and an external electric field, focusing on the resulting polarizability of the system as influenced by the coupling parameter. Finally, in Section~\ref{conclusions}, we present our conclusions, summarizing the key findings and emphasizing the significance of the effects of coupled systems in planar quantum systems.

\section{General Model}
\label{secII}

The following general Lagrangian density rigorously outlines the theoretical framework proposed herein:
\begin{align} 
\mathcal{L}&=\bar{\psi}\left( \frac{i\hbar}{2}\gamma ^{\mu }\overleftrightarrow{D}_{\mu }-m_{\psi} c^2\,I_{2}+\frac{g}{2}F^{\mu \nu }\sigma _{\mu \nu }\right) \psi\notag\\
&-\frac{1}{4\mu_0}F^{\mu \nu }F_{\mu \nu },
\end{align}
where $\psi$ represents a two-component spinor, $m_\psi$ is the particle's mass, and $g$ is the nonminimal coupling, which assumes the general form
\begin{equation}
g=\sum_{d=0}^{\infty }g^{\alpha _{1}\cdots \alpha_{d}}\partial _{\alpha _{1}}\ldots \partial _{\alpha _{d}}.
\end{equation}
The parameter $d$ dictates both the number of additional derivatives and the rank of the tensor $g^{\alpha _{1}\cdots \alpha_{d}}$, which regulates the anisotropies within the system. Moreover, it is associated with Lorentz-symmetry breaking in the planar photon sector \cite{ferreira2019dimensional,lisboa2023planar,ferrari2024interactions}.

The minimal interaction with the electromagnetic field is attained through the implementation of the covariant derivative $D_{\mu }=\partial _{\mu }-ieA_{\mu }$, whereas the nonminimal interaction term is characterized by $(g/2)F^{\mu \nu}\sigma _{\mu \nu }$, which can be articulated as $g\tilde{F}^{\lambda}\gamma_{\lambda}$, wherein the dual of $F^{\mu \nu }$ has been employed, i.e.,
$\tilde{F}^{\lambda}\equiv (1/2)F_{\mu \nu }\varepsilon^{\mu \nu
\lambda}$. Because the $\sigma_{\mu \nu}$ term is not an independent gamma matrix, the direct coupling of $F^{\mu \nu}$ to the fermionic sector is achieved only through the dual field. The algebra of the gamma matrices in $(1+2)$ dimensions is given by
\begin{align}
&\sigma_{\mu \nu} =\frac{i}{2}\left[ \gamma_{\mu },\gamma_{\nu}\right]
=\varepsilon_{\mu \nu \lambda}\gamma^{\lambda}, \\
&\left[ \gamma _{\mu },\gamma _{\nu }\right]  =-2i\varepsilon _{\mu \nu\lambda }\gamma ^{\lambda}, \\
&\left\{ \gamma _{\mu },\gamma _{\nu }\right\} =2\eta _{\mu \nu }I_{2}.
\end{align}

Since we intend to write the Dirac equation and subsequently access its non-relativistic limit, these results are explicitly evaluated in the following representation of the matrices:
\begin{equation}
\gamma ^{0}=
\begin{pmatrix}
1 & 0 \\ 
0 & -1
\end{pmatrix},\quad
\gamma ^{1}=
\begin{pmatrix}
0 & i \\ 
i & 0
\end{pmatrix},\quad
\gamma ^{2}=
\begin{pmatrix}
0 & 1 \\ 
-1 & 0
\end{pmatrix}.\label{dm}
\end{equation}

Following the introduction of the model, we now focus on a specific
case: the Lorentz-invariant scenario, which is obtained by setting $d = 0$. In this instance, the nonminimal coupling involves a standard electron interacting with the electromagnetic field within the framework of the Dirac equation
\begin{equation}
i\hbar \gamma ^{\mu }D_{\mu }\psi +g\tilde{F}^{\lambda }\gamma_{\lambda}\psi -m_{\psi} c^2\,I_{2}\psi=0.
\end{equation}
Using representation \eqref{dm} and decomposing the fermion field as
\begin{equation}
\begin{pmatrix}
\varphi  \\ 
\chi 
\end{pmatrix}
=
\begin{pmatrix}
\tilde{\varphi} \\ 
\tilde{\chi}
\end{pmatrix}
e^{-i m_{\psi} c^2 t/\hbar},
\end{equation}
we obtain the coupled equations
\begin{align}
i\hbar\partial _{0}\tilde{\varphi}& =\mathcal{D}_{\tilde{\chi}}\tilde{\chi}
-\left( eA_{0}+g\tilde{F}^{0}\right) \tilde{\varphi}, \label{O1}\\[1mm]
i\hbar\partial _{0}\tilde{\chi}& =\mathcal{D}_{\tilde{\varphi}}\tilde{\varphi}
-\left( 2m_{\psi} c^2+eA_{0}+g\tilde{F}^{0}\right) \tilde{\chi}, \label{O2}
\end{align}
where we have defined the operators $\mathcal{D}_{\tilde{\chi}}=-D^{1}+iD^{2}+ig\tilde{F}^{1}+g\tilde{F}^{2}$ and $\mathcal{D}_{\tilde{\varphi}}=D^{1}+iD^{2}-ig\tilde{F}^{1}+g\tilde{F}^{2}$.
The second-order equation implied by Eqs. (\ref{O1}) and (\ref{O2}) is obtained by applying the non-relativistic approximations: $| i\partial _{0}\tilde{\chi}| \ll
| m_{\psi }\tilde{\chi}| $, $| eA_{0}\tilde{\chi}
| \ll | m_{\psi }\tilde{\chi}| $, and $
| g\tilde{F}^{0}\tilde{\chi}| \ll | m_{\psi }
\tilde{\chi}|$. Thus, we get
\begin{equation}
\tilde{\chi}=\frac{1}{2m_{\psi} c^2}\left( 1-\xi +\xi ^{2}+\ldots \right) 
\mathcal{D}_{\tilde{\varphi}}\tilde{\varphi},
\end{equation}
where
\begin{equation}
\xi =\frac{1}{2m_{\psi} c^2}\left( eA_{0}+g\tilde{F}^{0}\right).
\end{equation}

At any order, the effective equation for the large component $\tilde{\varphi}$ can be written as
\begin{align}
&i\hbar\partial _{0}\tilde{\varphi} =-\frac{\mathcal{G}_{n}\left( \xi\right) }{2m_{\psi} c^2}\hbar^2\boldsymbol{\nabla}^{2}\tilde{\varphi}+\frac{ie\mathcal{G}_{n}\left( \xi\right) }{2m_{\psi} c^2}\hbar\left( \boldsymbol{\nabla}\cdot \boldsymbol{A}+2\boldsymbol{A}\cdot \boldsymbol{\nabla}\right) \tilde{\varphi}\notag \\
&\quad +\frac{e^{2}\mathcal{G}_{n}\left( \xi\right) }{2m_{\psi} c^2}\boldsymbol{A}^{2}\tilde{\varphi}-\left( \frac{e\mathcal{G}_{n}\left( \xi\right) }{2m_{\psi} c^2}-g\right) B\tilde{\varphi}-eA_{0}\tilde{\varphi}\notag \\
&\quad +\frac{ig\mathcal{G}_{n}\left( \xi\right) }{2m_{\psi} c^2}\hbar\dot{B}\tilde{\varphi}-\frac{ig\mathcal{G}_{n}\left( \xi\right) }{2m_{\psi} c^2}\left( \boldsymbol{E}\times \hbar\boldsymbol{\nabla}\tilde{\varphi}+ie\boldsymbol{A}\times \boldsymbol{E}\tilde{\varphi}\right)\notag \\
&\quad +\frac{g\mathcal{G}_{n}\left( \xi\right) }{2m_{\psi} c^2}\hbar\mathbf{
\nabla }\cdot \boldsymbol{E}\tilde{\varphi}+\frac{g^{2}\mathcal{G}_{n}\left( \xi\right) }{2m_{\psi} c^2}\boldsymbol{E}^{2}\tilde{\varphi}\notag\\
&\quad +\frac{1}{2m_{\psi} c^2}\left[ -\hbar\partial ^{1}\mathcal{G}_{n}\left( \xi\right) +i\hbar\partial ^{2}\mathcal{G}_{n}\left( \xi\right) \right] 
\mathcal{D}_{\tilde{\varphi}}\tilde{\varphi}, \label{egSI}
\end{align}
with
\begin{equation}
\mathcal{G}_{n}\left( \xi \right) =\sum_{m=0}^{n}\left( -1\right) ^{m}\xi^{m}=1-\xi +\xi ^{2}+\ldots. \label{aprSI}
\end{equation}

Given the complexity of the equation of motion (\ref{egSI}), and since we intend to apply the model to already consolidated quantum systems, 
we intend for now to consider only the zeroth order approximation in the $
\xi $ parameter. Replacing the approximation (\ref{aprSI}) with the original equation, we obtain
\begin{equation}
i\hbar\partial _{0}\tilde{\varphi}=\frac{1}{2m_{\psi} c^2}\mathcal{D}_{\tilde{\chi}}
\mathcal{D}_{\tilde{\varphi}}\tilde{\varphi}-\left( eA_{0}+g\tilde{F}^{0}\right) \tilde{\varphi}.\label{svSI}
\end{equation}
The last term in this expression can be developed and written more explicitly. To simplify some calculation realizations, we define the quantities $\mathfrak{F}=ig\tilde{F}^{1}+g\tilde{F}^{2}$ and $\mathfrak{F}^{\ast }=-ig\tilde{F}^{1}+g\tilde{F}^{2}$. In this way, Eq. (\ref{svSI}) assumes the form 
\begin{align}
i\hbar\partial _{0}\tilde{\varphi}
= {} & -\frac{\hbar^2}{2m_{\psi} c^2}\left( \boldsymbol{\nabla}-i\frac{e}{\hbar}\boldsymbol{A}\right) ^{2}\tilde{\varphi}-\frac{eB}{2m_{\psi} c^2}\hbar\tilde{\varphi}-eA_{0}\tilde{\varphi}\notag \\
& -g\tilde{F}^{0}\tilde{\varphi}+\frac{ig}{2m_{\psi} c^2}\hbar\partial ^{j}\tilde{F}^{j}\tilde{\varphi}-\frac{g}{2m_{\psi} c^2}\varepsilon ^{ij}\hbar\partial ^{i}\tilde{F}^{j}\tilde{\varphi}\notag \\
& +\frac{ig}{m_{\psi} c^2}\tilde{F}^{j}(-i\hbar\partial ^{j}-eA^j)\tilde{\varphi}+\frac{g^{2}}{2m_{\psi} c^2}\tilde{F}^{j}\tilde{F}^{j}\tilde{\varphi}.
\end{align}
By using
\begin{align}
\tilde{F}^{0}& \equiv \frac{1}{2}\varepsilon ^{ij0}F_{ij}=-B,\\[1mm]
\tilde{F}^{j}& \equiv \frac{1}{2}\varepsilon ^{\mu \nu j}F_{\mu \nu}=-\varepsilon ^{jm}E^{m},
\end{align}
and introducing the two-dimensional notation $\boldsymbol{x}\times \boldsymbol{y}=\varepsilon ^{jm}x^{j}y^{m}$, we obtain
\begin{align}
&i\hbar\partial _{0}\tilde{\varphi} =  -\frac{\hbar^2}{2m_{\psi} c^2}\nabla^2\tilde{\varphi}+\frac{i e\hbar}{2m_{\psi} c^2}\left( \nabla\cdot \boldsymbol{A}+2\boldsymbol{A}\cdot \nabla\right)\tilde{\varphi}\notag\\&+\frac{e^2}{2m_{\psi} c^2}\boldsymbol{A}^{2}\tilde{\varphi}-\left(\frac{e}{2m_{\psi} c^2}-g\right)B\tilde{\varphi}-eA_{0}\tilde{\varphi}\notag \\
& -\frac{ig}{2m_{\psi} c^2}\left( \boldsymbol{E}\times \nabla\tilde{\varphi}+i\frac{e}{\hbar}\boldsymbol{A}\times \boldsymbol{E}\tilde{\varphi}\right)\notag \\
& +\frac{ig}{2m_{\psi} c^2}\hbar\dot{B}\tilde{\varphi}+\frac{g}{2m_{\psi} c^2}\hbar(\nabla\cdot \boldsymbol{E})\tilde{\varphi}+\frac{g^{2}}{2m_{\psi} c^2}\boldsymbol{E}^{2}\tilde{\varphi},\label{order1SI}
\end{align}
with $\boldsymbol{\nabla}\times \boldsymbol{E}=-\dot{B}$ being the analogous Faraday law for the planar system.

Equation (\ref{order1SI}) offers a framework for investigating various quantum systems, particularly within the domains of CMP. This formulation describes the motion of a charged particle, such as an electron, subjected to both electric and magnetic fields while incorporating nonminimal couplings. Additionally, the inclusion of nonminimal couplings allows for the exploration of systems where spin-orbit interaction is significant, such as in quantum dots and spintronics applications, where control over electron spins is crucial for developing quantum technologies \cite{liu2016quantum}. Moreover, the higher-order interactions between electromagnetic fields and the particle, introduced by the non-minimal coupling term, can be applied to study non-linear optical responses in various materials. This opens the door to investigating phenomena such as harmonic generation and multi-photon processes \cite{adamo2009light,chang2006quantum,mckay2024spatially,barontini2022measuring,arias2012wispy,cresini2019towards,reidler2008quantum}. Furthermore, the form of the equation is also pertinent to the study of topological insulators, where the interplay between electric and magnetic fields and their influence on edge states is essential for understanding the robustness of these electronic systems against disorder \cite{xia2009large}. Finally, the relativistic nature of the equation makes it applicable to Dirac and Weyl semimetals, where electrons exhibit relativistic-like behavior. In these materials, the equation can be used to explore phenomena such as the chiral anomaly, which occurs when electric and magnetic fields are applied simultaneously \cite{burkov2015chiral,li2016chiral,potter2014quantum,huang2015observation,PhysRevB.104.075202}. 

Overall, Eq. (\ref{order1SI}) is highly suitable for studying the quantum Hall effect, where the coupling between the particle’s motion and the magnetic field leads to the characteristic quantized Hall conductance \cite{kane2005graphene,sitte2012topological}. This phenomenon, typically observed in two-dimensional electron systems, plays a crucial role in understanding topological states of matter. We will focus on this topic throughout Section \ref{HallSec}. Furthermore, in Section \ref{StarkSec} and also using Eq. (\ref{order1SI}), we investigate the Stark effect, which describes how energy levels in a quantum system shift due to the presence of an electric field.

\section{Modified Landau Quantization and Hall Conductivity}
\label{HallSec}

In this section, we apply Eq. (\ref{order1SI}) to the particular case where the particle moves in a uniform magnetic field to obtain the modified Landau levels and study the physical implications for Hall conductivity. This can be accomplished by making $A_{0}=0$, $\boldsymbol{E}=0$ and introducing the following configuration for the potential vector:
\begin{equation}
\boldsymbol{A}=\frac{1}{2}B\boldsymbol{r}_{\perp },\quad \boldsymbol{r}_{\perp}^{i}=\varepsilon^{ij}r^j,\quad \boldsymbol{r}=(x,y),\quad \boldsymbol{r}_{\perp}\cdot\boldsymbol{r}=0,\label{vectorA}
\end{equation}
which allows us to write Eq. (\ref{order1SI}) as
\begin{align}
i\hbar\partial _{0}\tilde{\varphi} = {} &
-\frac{\hbar^2}{2m_{\psi} c^2}\nabla^2\tilde{\varphi}+\frac{ie\hbar B}{2m_{\psi} c^2}\left(\boldsymbol{r}\times \nabla\right)\tilde{\varphi}+\frac{e^2B^2}{8m_{\psi} c^2}r^2\tilde{\varphi}\notag \\
&-\left(\frac{e}{2m_{\psi} c^2}-g\right)B\tilde{\varphi}.
\end{align}
For the field configuration in \eqref{vectorA}, the model describes the
Landau quantization problem renormalized by the coupling
constant $g$.
In analogy with the standard Landau quantization, the term $\left(e/2m_{\psi}c^2-g\right)B$ may be interpreted as a Zeeman-type term (with a correction due to the nonminimal coupling). Choosing wavefunctions of the form
\[
\tilde{\varphi}(r,\phi)=e^{-i\mathcal{E}t/\hbar}e^{im\phi}f(r),\quad m=0,\pm 1,\pm 2,\ldots,
\]
the radial equation is found to be
\begin{equation}
\left( \frac{d^{2}}{dr^{2}}+\frac{1}{r}\frac{d}{dr}-\frac{m^{2}}{r^{2}}
\right) f(r)-\frac{r^{2}}{4\lambda^4}f(r)+\varepsilon f(r)=0,
\end{equation}
where the magnetic length is
\[
\lambda=\sqrt{\frac{\hbar}{m_{\psi} c\, \omega_c}},
\]
the cyclotron frequency is
\[
\omega_c=\frac{eB}{m_{\psi} c},
\]
and
\[
\varepsilon=\frac{2m_{\psi} c^2}{\hbar}\left(\mathcal{E}-gB\right)+\frac{m_{\psi} c\, \omega_c}{\hbar}(1+m).
\]
The eigenvalues and eigenfunctions are, respectively,
\begin{equation}
\mathcal{E}_{nm}=\left[ n-m\Theta(-m)\right]\hbar\omega_c+gB,\label{eq:Energy-B}
\end{equation}
with $\Theta(x)$ the Heaviside step function, and
\begin{align}
\tilde{\varphi}_{nm}(r,\phi)= {} & \frac{1}{\lambda}\sqrt{\frac{\Gamma(n+|m|+1)}{2^{|m|+1}\left[\Gamma(|m|+1)\right]^2\pi}}\left(\frac{r}{\lambda}\right)^{|m|}\notag\\
&\times e^{im\phi}e^{-\frac{r^2}{4\lambda^2}}\,_1F_1\!\left(-n,|m|+1,\frac{r^2}{2\lambda^2}\right),
\end{align}
where $\Gamma(z)$ is the Gamma function and $_1F_1(a,b,z)$ is the confluent hypergeometric function.

For the numerical analysis presented in the figures below, we adopt natural units ($\hbar=1$, $m_{\psi}=1$, $c=1$).

\subsection{Hall Conductivity}
It is well established that, within the linear response approximation and when the Fermi energy lies inside the energy gap, the Hall conductivity at zero temperature is given by \cite{kolovsky2011hall,sharma2016nernst,ferreiros2017anomalous,markov2019robustness,wang2022quantum,ando2015valley}
\begin{equation}
\label{eq:Hall-1}
\sigma_H\left(\epsilon_F, 0\right) = \frac{e}{S} \frac{\partial N}{\partial B},
\end{equation}
where $ S $ is the surface area and $ N $ is the number of states below the Fermi energy $\epsilon_F$. The Density of States (DOS) per unit area, $D\left(\mathcal{E}\right)$, is expressed as \cite{ching1987solid,louis2002density}
\begin{equation}
\label{eq:DOS}
D\left(\mathcal{E}\right) = \frac{|eB|}{2\pi\hbar} \sum_{n,m} \delta\left(\mathcal{E} - \bar{\mathcal{E}}_{nm}\right),
\end{equation}
where $\mathcal{E}$ is the energy at which the density of states is being evaluated, and $\bar{\mathcal{E}}_{nm}$ is the energy of each particular state.

\begin{figure}[tbh]
\centering
\includegraphics[width=\columnwidth]{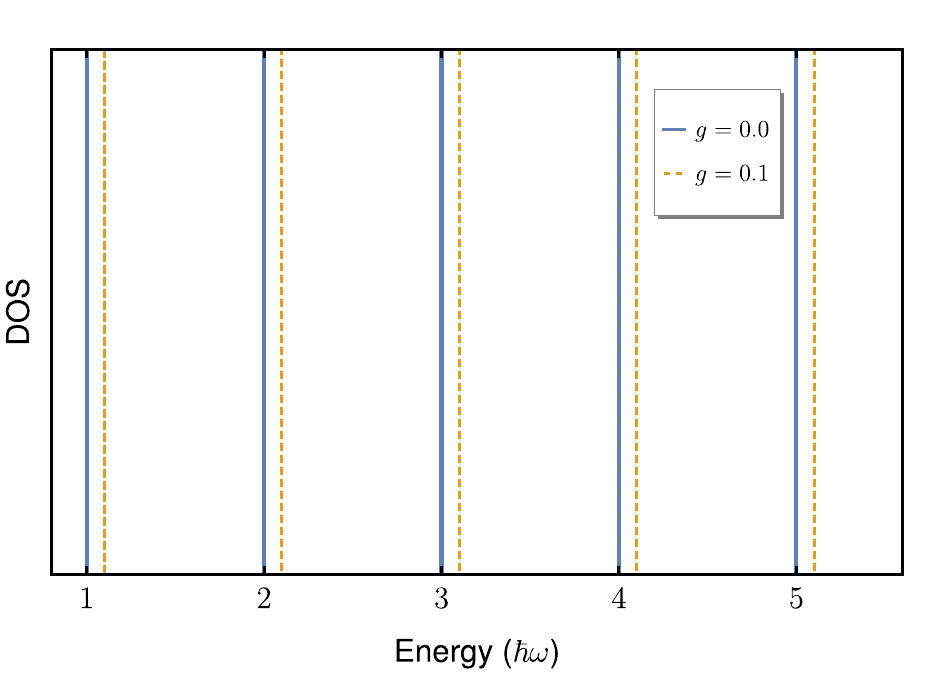}
\caption{\label{fig:DOS}
DOS given by Eq.~\eqref{eq:DOS} as a function of energy for two different values of $ g $ (fixed $ B = 1.0 $).}
\end{figure}

Figure~\ref{fig:DOS} schematically illustrates the discrete spectrum of the DOS using delta peaks. The parameter $ g $ shifts the DOS to higher energy values, indicating that the coupling parameter significantly influences the system's physical properties. For instance, it can affect the plateaus of conductivity.

Let us now analyze how this spectral shift due to $ g $ impacts the plateaus of conductivity. The total number of states below the Fermi energy $\epsilon_F$ is given by \cite{marder2010condensed,bruus2004manybody}
\begin{align}
\label{eq:numest}
N &= S \int_{-\infty}^{\epsilon_F} D\left(\mathcal{E}\right) d\mathcal{E} \notag \\
&= \frac{S |eB|}{2\pi\hbar} \left\lfloor \frac{m_{\psi} \epsilon_F}{eB} + m\Theta(-m) - g \right\rfloor,
\end{align}
where $\lfloor x \rfloor$ denotes the floor function.

Using Eq.~\eqref{eq:Hall-1} and noting that the integer part in Eq.~\eqref{eq:numest} remains constant for a given value of the Fermi energy, the Hall conductivity becomes
\begin{equation}
\frac{\sigma_H(\epsilon_F, 0)}{\sigma_0} = -\left\lfloor \frac{m_{\psi} \epsilon_F}{eB} + m\Theta(-m) - g \right\rfloor, \label{H1}
\end{equation}
where $\sigma_0 = e^2/h$ is the quantum of conductivity. From Eq.~\eqref{H1}, we observe that the Hall conductivity is quantized due to the floor function, exhibits a proportional relationship between the quantized plateaus and the Fermi energy $\epsilon_F$, and displays an inverse dependence on the magnetic field $B$. Furthermore, the coupling parameter $g$ significantly influences its value, shifting the quantization steps.

\begin{figure}[tbh]
\centering
\includegraphics[width=0.95\columnwidth]{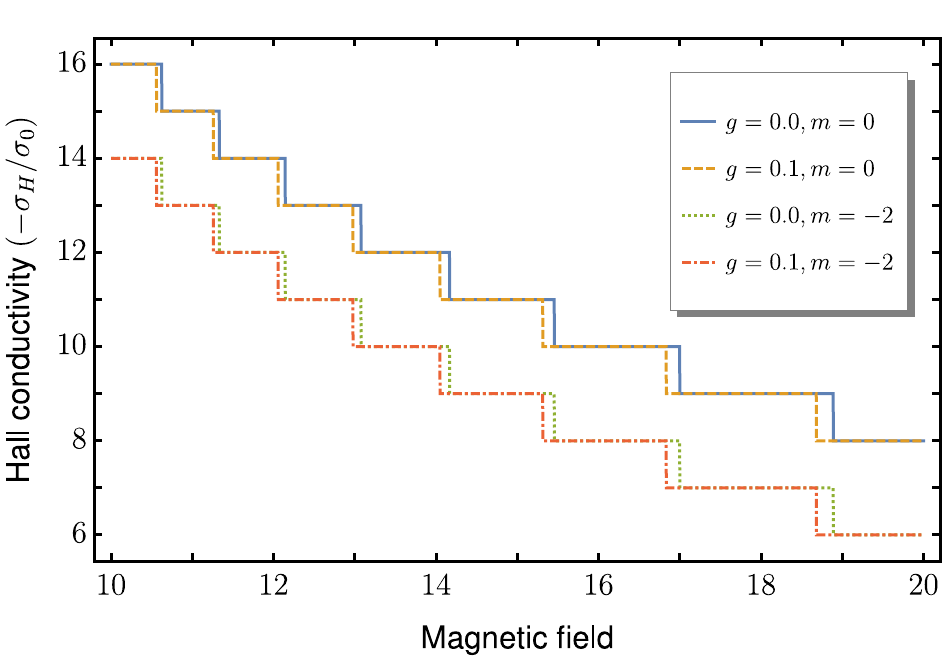}
\caption{
Hall conductivity at zero temperature as a function of the magnetic field for various values of $ g $ and $ m $. Fixed: $ m_{\psi}\epsilon_F/e = 170 $.}
\label{fig:fig2}
\end{figure}

\begin{figure}[tbh]
\centering
\includegraphics[width=0.95\columnwidth]{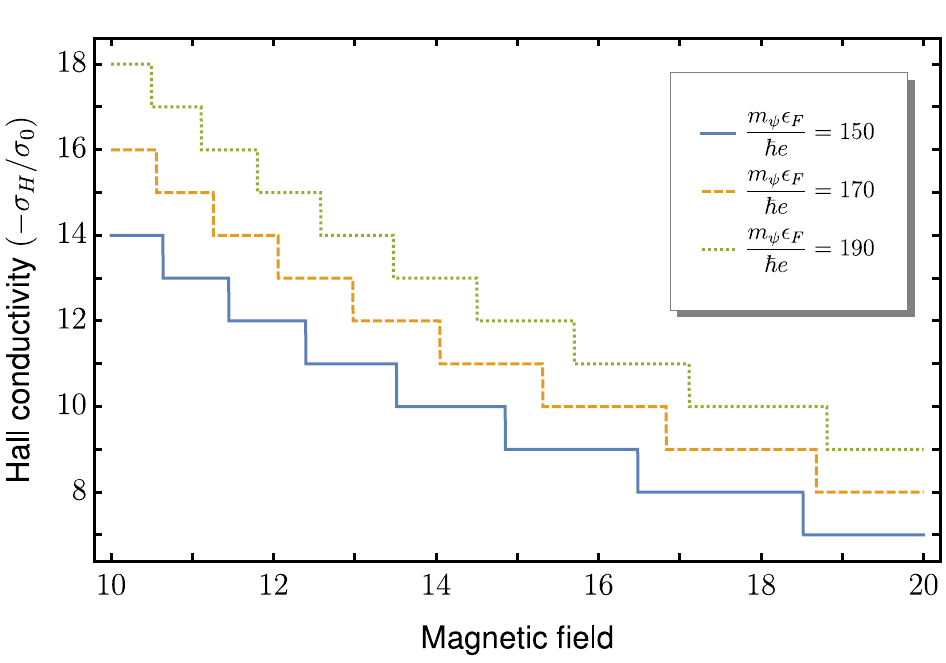}
\caption{
Hall conductivity at zero temperature as a function of the magnetic field for different Fermi energies $\epsilon_F$ with $ g = 0.1 $ and fixed $ m = 0 $.}
\label{fig:fig3}
\end{figure}

Figure~\ref{fig:fig2} shows the Hall conductivity at zero temperature as a function of $B$ for various values of $g$ and $m$ (with fixed $m_{\psi}\epsilon_F/e = 170$), while Fig.~\ref{fig:fig3} displays the results for different Fermi energies with $g = 0.1$ (and fixed $m = 0$). 

Figure~\ref{fig:fig2} illustrates that as $g$ increases, the plateaus shift to lower values of $B$. The plateau width increases with the magnetic field for fixed $g$ and $m$, and negative $m$ leads to smaller Hall conductivity values. In Fig.~\ref{fig:fig3}, with $g = 0.1$ and different values of $\epsilon_F$, it is observed that Landau levels approach each other as $\epsilon_F$ increases, resulting in narrower plateaus.

From the results of Figs.~\ref{fig:fig2} and \ref{fig:fig3}, it is evident that $ g $ significantly influences the Hall conductivity by altering the position and width of the plateaus. This demonstrates the impact of the system's quantization structure. In the
next section, we will continue to investigate the effect of $g$ in
other physical contexts, such as its relation to an applied electric
field.

\section{2D Harmonic Oscillator in an Electric Field and Stark Effect Corrections}
\label{StarkSec}

This section considers a 2D harmonic oscillator subjected to a uniform electric field. In addition, the system includes the coupling term $g$, which may introduce further modifications to the energy levels and affect the polarizability of the system. Studying quantum systems in external fields reveals significant perturbative effects on energy levels. One such phenomenon is the Quantum Stark Effect, which describes how energy levels shift due to the presence of an electric field \cite{cohen1977quantum,schiff1968quantum,demtroder2010atoms,jackson1998classical,tipler2012modern,friedrich2005theoretical,scully1997quantum}. 

We begin by considering the modified Schrödinger equation for a particle of mass $ m_{\psi} $ in the presence of a harmonic potential, an electric field along the $x$-axis, and the additional coupling term:
\begin{align}
i\hbar\partial_0 \tilde{\varphi} = & -\frac{\hbar^2}{2m_{\psi} c^2}\nabla^2\tilde{\varphi} + \frac{g^2 E_x^2}{2m_{\psi} c^2}\tilde{\varphi} + V_{HO}\tilde{\varphi} - eE_x x\,\tilde{\varphi}\notag \\
& - \frac{ig\hbar E_x}{2m_{\psi} c^2}\frac{\partial \tilde{\varphi}}{\partial y},
\end{align}
where $ \omega $ is the angular frequency of the harmonic oscillator, and $ E_x $ is the applied electric field. The potential $ V_{HO} = \frac{1}{2} m_{\psi} \omega^2 (x^2 + y^2) $ represents the unperturbed harmonic oscillator potential.

Assuming a stationary solution of the form
\[
\tilde{\varphi}(x, y, t) = \psi(x, y) e^{-i\mathcal{E}t/\hbar},
\]
we separate the equation into two independent differential equations for the coordinates $ x $ and $ y $, resulting in the wavefunction $ \psi(x, y) = \psi_x(x) \psi_y(y) $.

Hence, for the $y$-direction, the Schrödinger equation becomes
\begin{equation}
\mathcal{E}_y \psi_y(y) = -\frac{\hbar^2}{2m_{\psi}} \frac{d^2 \psi_y(y)}{dy^2} - \frac{ig \hbar E_x}{2m_{\psi}} \frac{d \psi_y(y)}{dy}.
\end{equation}
Introducing the shifted variable $ \tilde{y} = y + g \hbar E_x/m_{\psi} \omega^2$, we simplify this equation to the well-known harmonic oscillator form, leading to the energy levels
\[
\mathcal{E}_y = \left( n_y + \frac{1}{2} \right) \hbar \omega,
\]
and the corresponding wave functions
\[
\psi_y(\tilde{y}) = \left( \frac{m_{\psi} \omega}{\pi \hbar} \right)^{1/4} H_{n_y}\left( \sqrt{\frac{m_{\psi} \omega}{\hbar}} \tilde{y} \right) e^{-\frac{m_{\psi} \omega \tilde{y}^2}{2 \hbar}},
\]
where $ n_y $ is a nonnegative integer and $ H_{n_y} $ are the Hermite polynomials.

On the other hand, for the $ x $-direction, the Schrödinger equation is
\begin{align}
\mathcal{E}_x \psi_x(x) = {} & -\frac{\hbar^2}{2m_{\psi}} \frac{d^2 \psi_x(x)}{dx^2} + \frac{1}{2}m_{\psi}\omega^2 x^2 \psi_x(x) \notag \\
& - eE_x x \psi_x(x) + \frac{g^2 E_x^2}{2m_{\psi}} \psi_x(x).
\end{align}
Introducing the shifted variable $ x' = x - eE_x/m_{\psi}\omega^2$, we eliminate the linear term, yielding the standard harmonic oscillator equation
\begin{align}
\mathcal{E}_x \psi_x(x') = -&\frac{\hbar^2}{2m_{\psi}} \frac{d^2 \psi_x(x')}{dx'^2} + \frac{1}{2}m_{\psi}\omega^2 x'^2 \psi_x(x') \notag\\
&+ \frac{g^2 E_x^2}{2m_{\psi}} \psi_x(x'),
\end{align}
with the energy levels in the $ x $-direction being then
\[
\mathcal{E}_x = \left( n_x + \frac{1}{2} \right) \hbar \omega - \frac{e^2 E_x^2}{2m_{\psi} \omega^2} + \frac{g^2 E_x^2}{2m_{\psi}},
\]
and the wavefunctions are
\[
\psi_x(x') = \left( \frac{m_{\psi} \omega}{\pi \hbar} \right)^{1/4} H_{n_x}\left(\sqrt{\frac{m_{\psi} \omega}{\hbar}} x' \right) e^{-\frac{m_{\psi} \omega x'^2}{2 \hbar}},
\]
where $ n_x $ is also a non-negative integer.

The total energy of the system, considering both the $x$ and $y$-directions, is
\begin{equation}
\mathcal{E} = \mathcal{E}_x + \mathcal{E}_y = \left(2n + 1 \right) \hbar \omega - \frac{e^2 E_x^2}{2m_{\psi} \omega^2} + \frac{g^2 E_x^2}{2m_{\psi}},
\end{equation}
where we have assumed $n_x=n_y=n$.
This result shows the second-order corrections due to the electric field, represented by the term $-e^2 E_x^2/2m_{\psi} \omega^2$, and the additional positive shift $g^2 E_x^2/2m_{\psi}$ due to the coupling constant $g$.

\begin{figure}[t]
\centering
\includegraphics[width=0.45\textwidth]{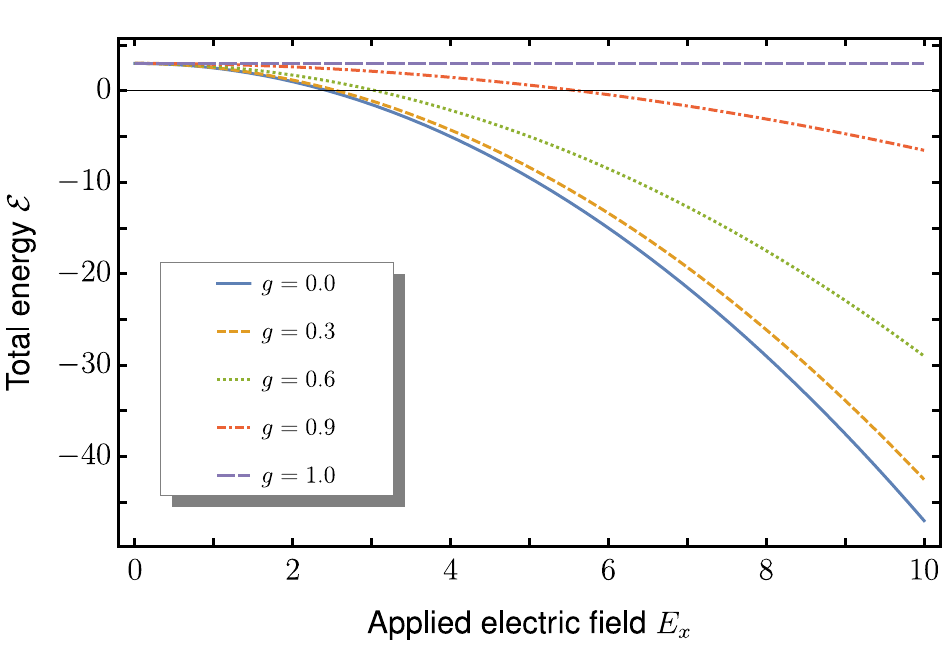}
\caption{The total energy as a function of the applied electric field for varying coupling parameter $g$. We use fixed $n=1$ and $\omega=1$.}
\label{fig:energy1}
\end{figure}

Figure \ref{fig:energy1} illustrates the degeneracy of energy levels under weak electric fields. As the coupling parameter increases, the energy curves become progressively more linear, eventually reaching exact linearity. For smaller values of $g$, the total energy decreases to negative values more rapidly as the electric field strength increases. The prolonged positivity of the energies is attributed to the additional positive shift term associated with the coupling parameter.

\begin{figure}[t]
\centering
\includegraphics[width=0.45\textwidth]{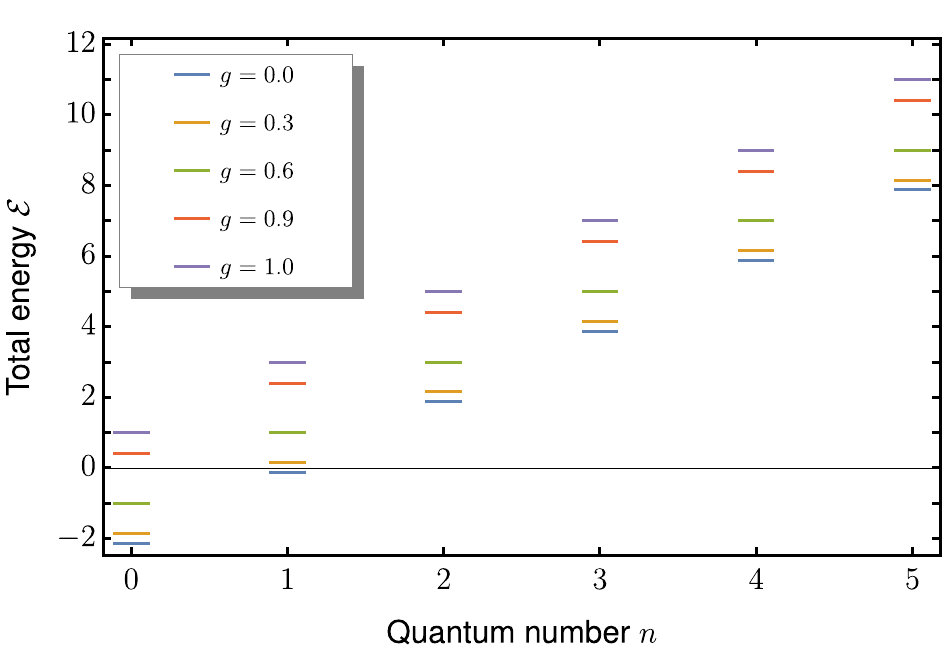}
\caption{The total energy as a function of the quantum number $n$ for varying coupling parameter $g$. We use fixed $E_x=2.5$ and $\omega=1$.}
\label{fig:energy2}
\end{figure}

Figure \ref{fig:energy2} also shows that energy levels become more positive when the coupling parameter increases. As expected, the energy levels become progressively more positive as the quantum number increases.

\begin{figure}[b]
\centering
\includegraphics[width=0.45\textwidth]{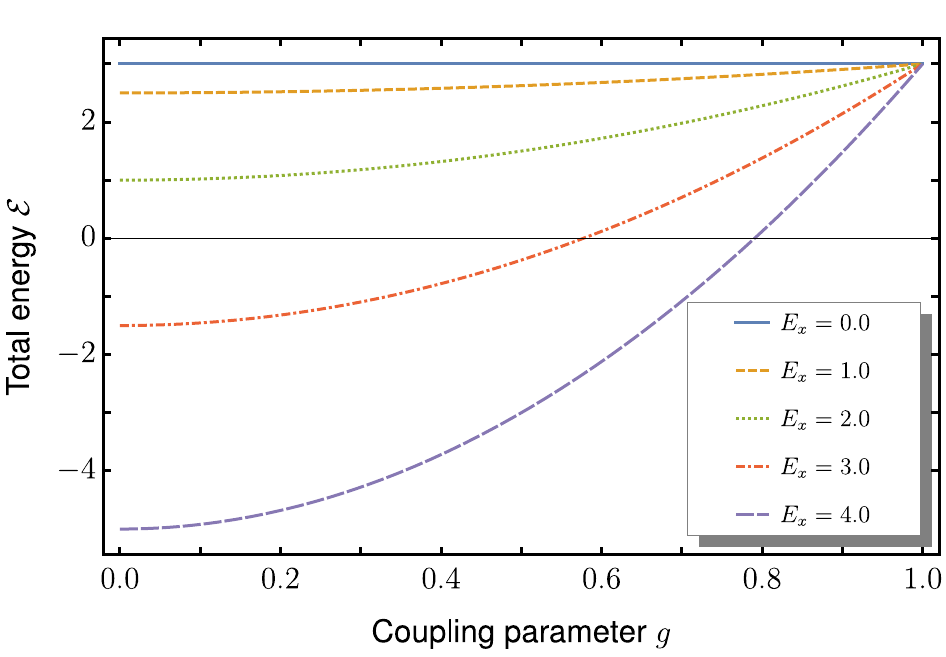}
\caption{The total energy as a function of the coupling parameter for varying applied electric field. We use $n=1$ and $\omega=1$.}
\label{fig:energy3}
\end{figure}

Figure \ref{fig:energy3} illustrates the degeneracy of energy levels for a coupling parameter of $g = 1$. As the applied electric field decreases, the energy curves become increasingly linear,  achieving exact linearity at $E_x = 0$. For lower values of $g$, the total energy is shifted to negative values more rapidly as the electric field strength increases. Note that $g$ is defined within the range $0 \leq g \leq 1$.

\begin{figure}[tbh]
\centering
\includegraphics[width=0.45\textwidth]{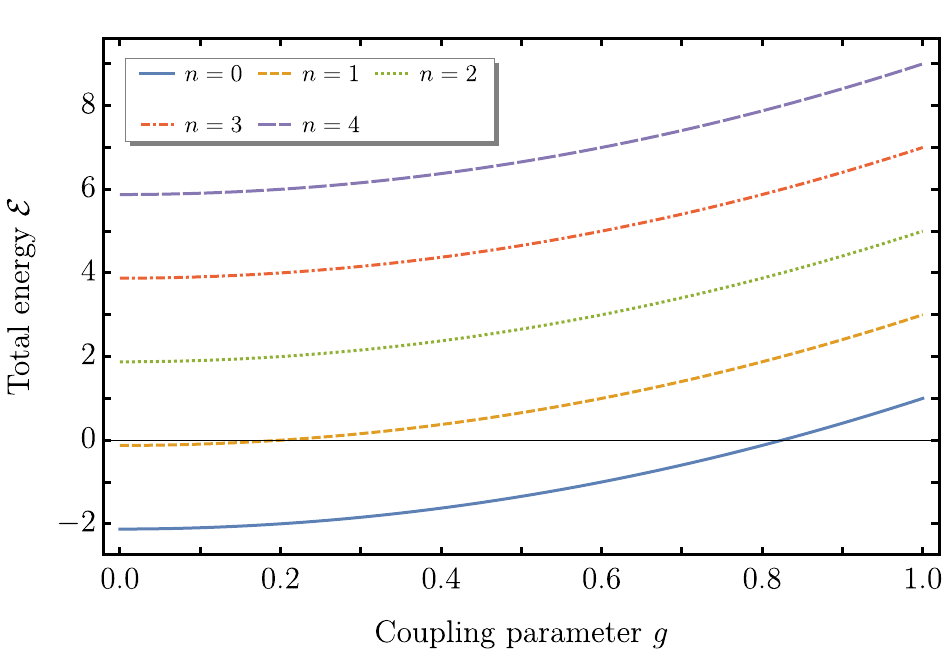}
\caption{The total energy as a function of the coupling parameter for varying quantum number $n$. Here, we also use fixed $E_x=2.5$ and $\omega=1$.}
\label{fig:energy4}
\end{figure}

Figure \ref{fig:energy4} shows that as the coupling parameter $g$ increases, the energy curves shift to more positive values. For lower values of $g$, the total energy decreases to negative values. Additionally, higher quantum numbers correspond to increasingly positive energy levels.
We emphasize that the coupling parameter is crucial in determining the energy levels. We will quantify the system's response to an applied electric field by calculating its polarizability to further characterize this effect.

\subsection{Effect of nonminimal coupling on the polarizability}

The polarizability $\alpha$ of the system describes its tendency to develop a dipole moment in response to an applied electric field. Mathematically, it is defined as the second derivative of the total energy with respect to the applied electric field $E_x$ \cite{VISHWAKARMA2023653}:
\begin{equation}
\alpha =-\frac{\partial^2 \mathcal{E}}{\partial E_x^2}= \frac{e^2}{m_{\psi} \omega^2} - \frac{g^2}{m_{\psi}}. \label{polarizability}
\end{equation}
The first term, in Eq. \eqref{polarizability}, represents the traditional polarizability due to the electric field, while the second term, influenced by the coupling constant $g$, reduces the overall polarizability. This shows how the system's response to external fields is modulated by the interaction term $g$.
\begin{figure}[b]
\centering
\includegraphics[width=0.45\textwidth]{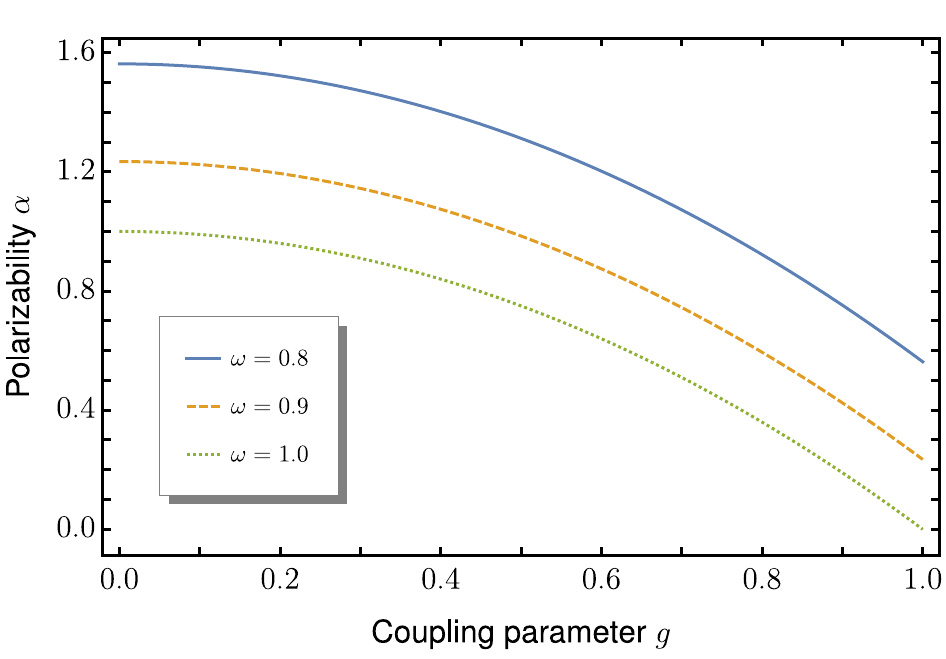}
\caption{Dependence of the polarizability $\alpha$ on the coupling parameter $g$ for varying values of frequency $\omega$.}
\label{fig:polarizability_g}
\end{figure}

For small values of $g$, such that $g \ll 1$ ($g^2 \ll e^2/\omega^2$), the polarizability is dominated by the traditional term $e^2/m_{\psi} \omega^2$. In this regime, the system responds to the electric field similarly to the standard 2D harmonic oscillator without nonminimal coupling. However, as the coupling parameter $g$ increases, the second term $g^2/m_{\psi}$ becomes more significant and can substantially reduce polarizability.
In the limit case where $g \approx 1$ ($g^2 \approx e^2/\omega^2$), the polarizability approaches zero, indicating that the system becomes nearly unresponsive to the external electric field. This phenomenon can be interpreted as a form of shielding where the nonminimal coupling counteracts the polarizing effect of the electric field.

In Figure~\ref{fig:polarizability_g}, to gain a deeper quantitative understanding of the expected behavior discussed above, we plot the polarizability $\alpha$ as a function of the coupling parameter $g$, within the range $0 \leq g \leq 1$ (in units of $e^2/\omega^2$, since we use $e = \omega = 1$).
The polarizability is higher for small values of $g$ and approaches the traditional value $e^2 / m_{\psi} \omega^2$. As $g$ increases, the polarizability decreases and eventually becomes zero (at least for $\omega = 1$). The point where $\alpha = 0$ indicates the value of $g$ at which the nonminimal coupling exactly cancels the traditional polarizability. Additionally, as the frequency $\omega$ increases, the energy values shift to lower values.
 This analysis demonstrates that the nonminimal coupling term $g$ significantly affects the system's polarizability and aligns with the expectations mentioned above.

For values of $g > 1$ (such that $g^2 > e^2/\omega^2$), the polarizability, as given by \eqref{polarizability}, becomes negative, which is an unusual behavior in classical terms. This phenomenon arises from the energy behavior within this range of $g$. A negative polarizability indicates that the system resists alignment with the electric field more strongly than in the absence of nonminimal coupling, potentially leading to counterintuitive effects, such as induced dipoles opposite to the applied field. In this work, we avoid choosing $g$ values exceeding 1. Instead, the range $0 \leq g \leq 1$ is more suitable for applications in CMP.

Before concluding this section, it is important to highlight that tuning $g$ has the potential to influence the system's response to external electric fields. This modulation may have significant practical implications for regulating quantum system behavior in applications such as quantum sensors or materials possessing tunable dielectric properties. Further experimental inquiries could examine the extent to which the coupling parameter $g$ is amenable to adjustment within physical systems and assess whether the anticipated suppression of polarizability is observable. Moreover, investigating alternative forms of coupling or considering higher-order corrections could yield more profound insights into the intricate dynamics between external fields and quantum systems.

\section{Conclusions \label{conclusions}}

In this paper, we investigated the dynamics of fermions in a (1+2)-dimensional planar electrodynamics framework, incorporating nonminimal coupling effects. Extending the Dirac equation with an additional interaction term, we explored novel modifications to the fermionic spectrum and their implications for condensed matter physics. Our results demonstrated that nonminimal coupling fundamentally alters key physical properties, such as the DOS, Hall conductivity, and energy levels in the presence of an external electric field.

The analysis of Landau quantization in the presence of nonminimal coupling revealed that the parameter $g$ induces a shift in the Landau levels, consequently modifying the Hall conductivity plateaus. The numerical results showed that increasing $g$ shifts the plateaus to lower magnetic field values while also affecting the plateau widths. This behavior suggests that nonminimal coupling provides a tunable mechanism for modifying transport properties in planar fermionic systems. The quantization structure of the Hall conductivity was directly influenced by the interplay between the coupling parameter, the Fermi energy, and the applied magnetic field, introducing new effects that may be experimentally accessible in condensed matter platforms.

Additionally, we investigated the impact of nonminimal coupling on a two-dimensional harmonic oscillator subject to an external electric field. The results showed that the coupling parameter modifies the Stark effect, introducing a shift in energy levels that depends on the external field strength. Notably, $g$ alters the system’s polarizability, effectively reducing its response to the applied electric field. This result highlights the possibility of engineering tunable quantum systems where the response to external perturbations can be modulated by adjusting the coupling parameter.

While our analysis focused on fundamental aspects of nonminimal coupling in planar fermionic systems, several open questions remain. One important aspect is the role of anisotropies induced by higher-order derivative terms in the coupling parameter $g$. Since anisotropic effects play a crucial role in condensed matter physics, further studies could investigate how directional dependence emerges in these systems and its potential impact on transport and optical properties. Moreover, a deeper exploration of quantum corrections and renormalization effects in the presence of nonminimal couplings could provide valuable insights into the stability of these models.

Another promising direction is the study of nonminimal couplings in the context of topological materials, such as graphene and topological insulators. The interplay between nonminimal interactions and topological effects may lead to new quantum phases or exotic transport phenomena. Additionally, investigating the coupling between planar fermionic systems and gravitational fields could extend the analysis to curved spacetimes, offering new perspectives on the AdS/CFT correspondence and potential quantum gravity applications.

In summary, this paper lays the groundwork for a broadmrange of future studies on the dynamics of fermions in ($1+2$)-dimensional systems, particularly concerning nonminimal couplings, Lorentz violations, and the anisotropies they induce. The study of planar fermions with nonminimal couplings is a rich and diverse field with far-reaching implications across multiple areas of physics. It provides fertile ground for exploring potential deviations from established symmetries, contributing to the broader quest for new physics beyond the SM. Continued research in this field is expected to deepen our understanding of quantum phenomena in lowerdimensional systems, advance our knowledge of fermionic interactions in two dimensions, and contribute to advancements in quantum materials and high-energy physics.

Finally, we will continue investigating the influence of nonminimal coupling on the physics of planar fermions in the context of Lorentz violations, with further advancements to be reported in future publications.

\section*{Acknowledgments}

This work was partially supported by the Brazilian agencies CAPES, CNPq, and FAPEMA. Edilberto O. Silva acknowledges the support from grants CNPq/306308/2022-3, FAPEMA/UNIVERSAL-06395/22, and FAPEMA/APP-12256/22. L. Lisboa-Santos is supported by Conselho Nacional de Desenvolvimento Cient\'{\i}fico e Tecnol\'ogico (CNPq/PDJ 151424/2022). F. S. Azevedo acknowledges CNPq Grant No. 153635/2024-0. This study was partly financed by the Coordena\c{c}\~ao de Aperfei\c{c}oamento de Pessoal de N\'{\i}vel Superior (CAPES) - Brazil (Code 001). C.D. FMA acknowledges financial support by CNPq grant 313124/2023-0.

\bibliographystyle{apsrev4-2}

\begin{thebibliography}{125}%
\makeatletter
\providecommand \@ifxundefined [1]{%
 \@ifx{#1\undefined}
}%
\providecommand \@ifnum [1]{%
 \ifnum #1\expandafter \@firstoftwo
 \else \expandafter \@secondoftwo
 \fi
}%
\providecommand \@ifx [1]{%
 \ifx #1\expandafter \@firstoftwo
 \else \expandafter \@secondoftwo
 \fi
}%
\providecommand \natexlab [1]{#1}%
\providecommand \enquote  [1]{``#1''}%
\providecommand \bibnamefont  [1]{#1}%
\providecommand \bibfnamefont [1]{#1}%
\providecommand \citenamefont [1]{#1}%
\providecommand \href@noop [0]{\@secondoftwo}%
\providecommand \href [0]{\begingroup \@sanitize@url \@href}%
\providecommand \@href[1]{\@@startlink{#1}\@@href}%
\providecommand \@@href[1]{\endgroup#1\@@endlink}%
\providecommand \@sanitize@url [0]{\catcode `\\12\catcode `\$12\catcode
  `\&12\catcode `\#12\catcode `\^12\catcode `\_12\catcode `\%12\relax}%
\providecommand \@@startlink[1]{}%
\providecommand \@@endlink[0]{}%
\providecommand \url  [0]{\begingroup\@sanitize@url \@url }%
\providecommand \@url [1]{\endgroup\@href {#1}{\urlprefix }}%
\providecommand \urlprefix  [0]{URL }%
\providecommand \Eprint [0]{\href }%
\providecommand \doibase [0]{https://doi.org/}%
\providecommand \selectlanguage [0]{\@gobble}%
\providecommand \bibinfo  [0]{\@secondoftwo}%
\providecommand \bibfield  [0]{\@secondoftwo}%
\providecommand \translation [1]{[#1]}%
\providecommand \BibitemOpen [0]{}%
\providecommand \bibitemStop [0]{}%
\providecommand \bibitemNoStop [0]{.\EOS\space}%
\providecommand \EOS [0]{\spacefactor3000\relax}%
\providecommand \BibitemShut  [1]{\csname bibitem#1\endcsname}%
\let\auto@bib@innerbib\@empty
\bibitem [{\citenamefont {Abergel}\ \emph {et~al.}(2010)\citenamefont
  {Abergel}, \citenamefont {Apalkov}, \citenamefont {Berashevich},
  \citenamefont {Ziegler},\ and\ \citenamefont {Chakraborty}}]{AP.2010.59.261}%
  \BibitemOpen
  \bibfield  {author} {\bibinfo {author} {\bibfnamefont {D.}~\bibnamefont
  {Abergel}}, \bibinfo {author} {\bibfnamefont {V.}~\bibnamefont {Apalkov}},
  \bibinfo {author} {\bibfnamefont {J.}~\bibnamefont {Berashevich}}, \bibinfo
  {author} {\bibfnamefont {K.}~\bibnamefont {Ziegler}},\ and\ \bibinfo {author}
  {\bibfnamefont {T.}~\bibnamefont {Chakraborty}},\ }\href
  {https://doi.org/10.1080/00018732.2010.487978} {\bibfield  {journal}
  {\bibinfo  {journal} {Advances in Physics}\ }\textbf {\bibinfo {volume}
  {59}},\ \bibinfo {pages} {261} (\bibinfo {year} {2010})}\BibitemShut
  {NoStop}%
\bibitem [{\citenamefont {Gan}\ \emph {et~al.}(2022)\citenamefont {Gan},
  \citenamefont {Englund}, \citenamefont {Van~Thourhout},\ and\ \citenamefont
  {Zhao}}]{APR.2022.9.021302}%
  \BibitemOpen
  \bibfield  {author} {\bibinfo {author} {\bibfnamefont {X.}~\bibnamefont
  {Gan}}, \bibinfo {author} {\bibfnamefont {D.}~\bibnamefont {Englund}},
  \bibinfo {author} {\bibfnamefont {D.}~\bibnamefont {Van~Thourhout}},\ and\
  \bibinfo {author} {\bibfnamefont {J.}~\bibnamefont {Zhao}},\ }\href
  {https://doi.org/10.1063/5.0078416} {\bibfield  {journal} {\bibinfo
  {journal} {Applied Physics Reviews}\ }\textbf {\bibinfo {volume} {9}},\
  \bibinfo {pages} {021302} (\bibinfo {year} {2022})}\BibitemShut {NoStop}%
\bibitem [{\citenamefont {Piquero-Zulaica}\ \emph {et~al.}(2022)\citenamefont
  {Piquero-Zulaica}, \citenamefont {Lobo-Checa}, \citenamefont {El-Fattah},
  \citenamefont {Ortega}, \citenamefont {Klappenberger}, \citenamefont
  {Auw\"arter},\ and\ \citenamefont {Barth}}]{RMP.2022.94.045008}%
  \BibitemOpen
  \bibfield  {author} {\bibinfo {author} {\bibfnamefont {I.}~\bibnamefont
  {Piquero-Zulaica}}, \bibinfo {author} {\bibfnamefont {J.}~\bibnamefont
  {Lobo-Checa}}, \bibinfo {author} {\bibfnamefont {Z.~M.~A.}\ \bibnamefont
  {El-Fattah}}, \bibinfo {author} {\bibfnamefont {J.~E.}\ \bibnamefont
  {Ortega}}, \bibinfo {author} {\bibfnamefont {F.}~\bibnamefont
  {Klappenberger}}, \bibinfo {author} {\bibfnamefont {W.}~\bibnamefont
  {Auw\"arter}},\ and\ \bibinfo {author} {\bibfnamefont {J.~V.}\ \bibnamefont
  {Barth}},\ }\href {https://doi.org/10.1103/RevModPhys.94.045008} {\bibfield
  {journal} {\bibinfo  {journal} {Rev. Mod. Phys.}\ }\textbf {\bibinfo {volume}
  {94}},\ \bibinfo {pages} {045008} (\bibinfo {year} {2022})}\BibitemShut
  {NoStop}%
\bibitem [{\citenamefont {Hua}\ \emph {et~al.}(2024)\citenamefont {Hua},
  \citenamefont {Wang}, \citenamefont {Zhu},\ and\ \citenamefont
  {Chen}}]{PRB.2024.109.115429}%
  \BibitemOpen
  \bibfield  {author} {\bibinfo {author} {\bibfnamefont {J.}~\bibnamefont
  {Hua}}, \bibinfo {author} {\bibfnamefont {Z.~F.}\ \bibnamefont {Wang}},
  \bibinfo {author} {\bibfnamefont {W.}~\bibnamefont {Zhu}},\ and\ \bibinfo
  {author} {\bibfnamefont {W.}~\bibnamefont {Chen}},\ }\href
  {https://doi.org/10.1103/PhysRevB.109.115429} {\bibfield  {journal} {\bibinfo
   {journal} {Phys. Rev. B}\ }\textbf {\bibinfo {volume} {109}},\ \bibinfo
  {pages} {115429} (\bibinfo {year} {2024})}\BibitemShut {NoStop}%
\bibitem [{\citenamefont {Skiff}\ \emph {et~al.}(2023)\citenamefont {Skiff},
  \citenamefont {de~Juan}, \citenamefont {Queiroz}, \citenamefont {Mathimalar},
  \citenamefont {Beidenkopf},\ and\ \citenamefont {Ilan}}]{SPPC.2023.6.011}%
  \BibitemOpen
  \bibfield  {author} {\bibinfo {author} {\bibfnamefont {R.~M.}\ \bibnamefont
  {Skiff}}, \bibinfo {author} {\bibfnamefont {F.}~\bibnamefont {de~Juan}},
  \bibinfo {author} {\bibfnamefont {R.}~\bibnamefont {Queiroz}}, \bibinfo
  {author} {\bibfnamefont {S.}~\bibnamefont {Mathimalar}}, \bibinfo {author}
  {\bibfnamefont {H.}~\bibnamefont {Beidenkopf}},\ and\ \bibinfo {author}
  {\bibfnamefont {R.}~\bibnamefont {Ilan}},\ }\href
  {https://doi.org/10.21468/SciPostPhysCore.6.1.011} {\bibfield  {journal}
  {\bibinfo  {journal} {SciPost Phys. Core}\ }\textbf {\bibinfo {volume} {6}},\
  \bibinfo {pages} {011} (\bibinfo {year} {2023})}\BibitemShut {NoStop}%
\bibitem [{\citenamefont {Read}\ and\ \citenamefont
  {Green}(2000)}]{PRB.2000.61.10267}%
  \BibitemOpen
  \bibfield  {author} {\bibinfo {author} {\bibfnamefont {N.}~\bibnamefont
  {Read}}\ and\ \bibinfo {author} {\bibfnamefont {D.}~\bibnamefont {Green}},\
  }\href {https://doi.org/10.1103/PhysRevB.61.10267} {\bibfield  {journal}
  {\bibinfo  {journal} {Phys. Rev. B}\ }\textbf {\bibinfo {volume} {61}},\
  \bibinfo {pages} {10267} (\bibinfo {year} {2000})}\BibitemShut {NoStop}%
\bibitem [{\citenamefont {Yanase}\ \emph {et~al.}(2022)\citenamefont {Yanase},
  \citenamefont {Daido}, \citenamefont {Takasan},\ and\ \citenamefont
  {Yoshida}}]{PE.2022.140.115143}%
  \BibitemOpen
  \bibfield  {author} {\bibinfo {author} {\bibfnamefont {Y.}~\bibnamefont
  {Yanase}}, \bibinfo {author} {\bibfnamefont {A.}~\bibnamefont {Daido}},
  \bibinfo {author} {\bibfnamefont {K.}~\bibnamefont {Takasan}},\ and\ \bibinfo
  {author} {\bibfnamefont {T.}~\bibnamefont {Yoshida}},\ }\href
  {https://doi.org/https://doi.org/10.1016/j.physe.2022.115143} {\bibfield
  {journal} {\bibinfo  {journal} {Physica E: Low-dimensional Systems and
  Nanostructures}\ }\textbf {\bibinfo {volume} {140}},\ \bibinfo {pages}
  {115143} (\bibinfo {year} {2022})}\BibitemShut {NoStop}%
\bibitem [{\citenamefont {Feldman}\ and\ \citenamefont
  {Salmhofer}(2008)}]{RMP.2008.20.275}%
  \BibitemOpen
  \bibfield  {author} {\bibinfo {author} {\bibfnamefont {J.}~\bibnamefont
  {Feldman}}\ and\ \bibinfo {author} {\bibfnamefont {M.}~\bibnamefont
  {Salmhofer}},\ }\href {https://doi.org/10.1142/S0129055X08003304} {\bibfield
  {journal} {\bibinfo  {journal} {Reviews in Mathematical Physics}\ }\textbf
  {\bibinfo {volume} {20}},\ \bibinfo {pages} {275} (\bibinfo {year}
  {2008})}\BibitemShut {NoStop}%
\bibitem [{\citenamefont {Kopciuszy\'nski}\ \emph {et~al.}(2024)\citenamefont
  {Kopciuszy\'nski}, \citenamefont {Stepniak-Dybala}, \citenamefont {Zdyb},\
  and\ \citenamefont {Krawiec}}]{NL.2024.24.2175}%
  \BibitemOpen
  \bibfield  {author} {\bibinfo {author} {\bibfnamefont {M.}~\bibnamefont
  {Kopciuszy\'nski}}, \bibinfo {author} {\bibfnamefont {A.}~\bibnamefont
  {Stepniak-Dybala}}, \bibinfo {author} {\bibfnamefont {R.}~\bibnamefont
  {Zdyb}},\ and\ \bibinfo {author} {\bibfnamefont {M.}~\bibnamefont
  {Krawiec}},\ }\href {https://doi.org/10.1021/acs.nanolett.3c04046} {\bibfield
   {journal} {\bibinfo  {journal} {Nano Letters}\ }\textbf {\bibinfo {volume}
  {24}},\ \bibinfo {pages} {2175} (\bibinfo {year} {2024})}\BibitemShut
  {NoStop}%
\bibitem [{\citenamefont {Grumiller}\ and\ \citenamefont
  {Meyer}(2006)}]{CQG.2006.23.6435}%
  \BibitemOpen
  \bibfield  {author} {\bibinfo {author} {\bibfnamefont {D.}~\bibnamefont
  {Grumiller}}\ and\ \bibinfo {author} {\bibfnamefont {R.}~\bibnamefont
  {Meyer}},\ }\href {https://doi.org/10.1088/0264-9381/23/22/021} {\bibfield
  {journal} {\bibinfo  {journal} {Classical and Quantum Gravity}\ }\textbf
  {\bibinfo {volume} {23}},\ \bibinfo {pages} {6435} (\bibinfo {year}
  {2006})}\BibitemShut {NoStop}%
\bibitem [{\citenamefont {Grignani}\ \emph {et~al.}(1996)\citenamefont
  {Grignani}, \citenamefont {Plyushchay},\ and\ \citenamefont
  {Sodano}}]{NPB.1996.464.189}%
  \BibitemOpen
  \bibfield  {author} {\bibinfo {author} {\bibfnamefont {G.}~\bibnamefont
  {Grignani}}, \bibinfo {author} {\bibfnamefont {M.}~\bibnamefont
  {Plyushchay}},\ and\ \bibinfo {author} {\bibfnamefont {P.}~\bibnamefont
  {Sodano}},\ }\href
  {https://doi.org/https://doi.org/10.1016/0550-3213(96)00062-4} {\bibfield
  {journal} {\bibinfo  {journal} {Nuclear Physics B}\ }\textbf {\bibinfo
  {volume} {464}},\ \bibinfo {pages} {189} (\bibinfo {year}
  {1996})}\BibitemShut {NoStop}%
\bibitem [{\citenamefont {Cabra}\ and\ \citenamefont
  {Rossini}(2002)}]{EPJB.2002.29.529}%
  \BibitemOpen
  \bibfield  {author} {\bibinfo {author} {\bibfnamefont {D.~C.}\ \bibnamefont
  {Cabra}}\ and\ \bibinfo {author} {\bibfnamefont {G.~L.}\ \bibnamefont
  {Rossini}},\ }\href {https://doi.org/10.1140/epjb/e2002-00318-3} {\bibfield
  {journal} {\bibinfo  {journal} {The European Physical Journal B - Condensed
  Matter and Complex Systems}\ }\textbf {\bibinfo {volume} {29}},\ \bibinfo
  {pages} {529} (\bibinfo {year} {2002})}\BibitemShut {NoStop}%
\bibitem [{\citenamefont {Nguyen}\ and\ \citenamefont
  {Nguyen}(2016)}]{ANS.2016.7.025003}%
  \BibitemOpen
  \bibfield  {author} {\bibinfo {author} {\bibfnamefont {B.~H.}\ \bibnamefont
  {Nguyen}}\ and\ \bibinfo {author} {\bibfnamefont {V.~H.}\ \bibnamefont
  {Nguyen}},\ }\href {https://doi.org/10.1088/2043-6262/7/2/025003} {\bibfield
  {journal} {\bibinfo  {journal} {Advances in Natural Sciences: Nanoscience and
  Nanotechnology}\ }\textbf {\bibinfo {volume} {7}},\ \bibinfo {pages} {025003}
  (\bibinfo {year} {2016})}\BibitemShut {NoStop}%
\bibitem [{\citenamefont {Sakita}(1994)}]{Book.Sakita.146}%
  \BibitemOpen
  \bibfield  {author} {\bibinfo {author} {\bibfnamefont {B.}~\bibnamefont
  {Sakita}},\ }in\ \href@noop {} {\emph {\bibinfo {booktitle} {Correlation
  Effects in Low-Dimensional Electron Systems}}},\ \bibinfo {editor} {edited
  by\ \bibinfo {editor} {\bibfnamefont {A.}~\bibnamefont {Okiji}}\ and\
  \bibinfo {editor} {\bibfnamefont {N.}~\bibnamefont {Kawakami}}}\ (\bibinfo
  {publisher} {Springer Berlin Heidelberg},\ \bibinfo {address} {Berlin,
  Heidelberg},\ \bibinfo {year} {1994})\ pp.\ \bibinfo {pages}
  {146--153}\BibitemShut {NoStop}%
\bibitem [{\citenamefont {Karowski}\ \emph {et~al.}(1985)\citenamefont
  {Karowski}, \citenamefont {Schrader},\ and\ \citenamefont
  {Thun}}]{CMP.1985.97.5}%
  \BibitemOpen
  \bibfield  {author} {\bibinfo {author} {\bibfnamefont {M.}~\bibnamefont
  {Karowski}}, \bibinfo {author} {\bibfnamefont {R.}~\bibnamefont {Schrader}},\
  and\ \bibinfo {author} {\bibfnamefont {H.~J.}\ \bibnamefont {Thun}},\ }\href
  {https://doi.org/https://doi.org/10.1007/BF01206176} {\bibfield  {journal}
  {\bibinfo  {journal} {Communications in Mathematical Physics}\ }\textbf
  {\bibinfo {volume} {97}},\ \bibinfo {pages} {5 } (\bibinfo {year}
  {1985})}\BibitemShut {NoStop}%
\bibitem [{\citenamefont {Hatsugai}\ \emph {et~al.}(1996)\citenamefont
  {Hatsugai}, \citenamefont {Kohmoto},\ and\ \citenamefont
  {Wu}}]{PRB.1996.54.4898}%
  \BibitemOpen
  \bibfield  {author} {\bibinfo {author} {\bibfnamefont {Y.}~\bibnamefont
  {Hatsugai}}, \bibinfo {author} {\bibfnamefont {M.}~\bibnamefont {Kohmoto}},\
  and\ \bibinfo {author} {\bibfnamefont {Y.-S.}\ \bibnamefont {Wu}},\ }\href
  {https://doi.org/10.1103/PhysRevB.54.4898} {\bibfield  {journal} {\bibinfo
  {journal} {Phys. Rev. B}\ }\textbf {\bibinfo {volume} {54}},\ \bibinfo
  {pages} {4898} (\bibinfo {year} {1996})}\BibitemShut {NoStop}%
\bibitem [{\citenamefont {Sharon}\ \emph {et~al.}(2015)\citenamefont {Sharon},
  \citenamefont {Sharon}, \citenamefont {Shinohara},\ and\ \citenamefont
  {Tiwari}}]{sharon2015graphene}%
  \BibitemOpen
  \bibfield  {author} {\bibinfo {author} {\bibfnamefont {M.}~\bibnamefont
  {Sharon}}, \bibinfo {author} {\bibfnamefont {M.}~\bibnamefont {Sharon}},
  \bibinfo {author} {\bibfnamefont {H.}~\bibnamefont {Shinohara}},\ and\
  \bibinfo {author} {\bibfnamefont {A.}~\bibnamefont {Tiwari}},\ }\href@noop {}
  {\emph {\bibinfo {title} {Graphene: An Introduction to the Fundamentals and
  Industrial Applications}}},\ Advanced Material Series\ (\bibinfo  {publisher}
  {Wiley},\ \bibinfo {year} {2015})\BibitemShut {NoStop}%
\bibitem [{\citenamefont {Baker}\ and\ \citenamefont
  {Tallentire}(2022)}]{baker2022graphene}%
  \BibitemOpen
  \bibfield  {author} {\bibinfo {author} {\bibfnamefont {J.}~\bibnamefont
  {Baker}}\ and\ \bibinfo {author} {\bibfnamefont {J.}~\bibnamefont
  {Tallentire}},\ }\href@noop {} {\emph {\bibinfo {title} {Graphene: The Route
  to Commercialisation}}}\ (\bibinfo  {publisher} {Jenny Stanford Publishing},\
  \bibinfo {year} {2022})\BibitemShut {NoStop}%
\bibitem [{\citenamefont {Warner}\ \emph {et~al.}(2012)\citenamefont {Warner},
  \citenamefont {Schaffel}, \citenamefont {Rummeli},\ and\ \citenamefont
  {Bachmatiuk}}]{warner2012graphene}%
  \BibitemOpen
  \bibfield  {author} {\bibinfo {author} {\bibfnamefont {J.}~\bibnamefont
  {Warner}}, \bibinfo {author} {\bibfnamefont {F.}~\bibnamefont {Schaffel}},
  \bibinfo {author} {\bibfnamefont {M.}~\bibnamefont {Rummeli}},\ and\ \bibinfo
  {author} {\bibfnamefont {A.}~\bibnamefont {Bachmatiuk}},\ }\href@noop {}
  {\emph {\bibinfo {title} {Graphene: Fundamentals and Emergent
  Applications}}}\ (\bibinfo  {publisher} {Elsevier Science},\ \bibinfo {year}
  {2012})\BibitemShut {NoStop}%
\bibitem [{\citenamefont {Novoselov}\ \emph {et~al.}(2005)\citenamefont
  {Novoselov}, \citenamefont {Geim}, \citenamefont {Morozov}, \citenamefont
  {Jiang}, \citenamefont {Katsnelson}, \citenamefont {Grigorieva},
  \citenamefont {Dubonos},\ and\ \citenamefont {Firsov}}]{Nature.2005.438.197}%
  \BibitemOpen
  \bibfield  {author} {\bibinfo {author} {\bibfnamefont {K.~S.}\ \bibnamefont
  {Novoselov}}, \bibinfo {author} {\bibfnamefont {A.~K.}\ \bibnamefont {Geim}},
  \bibinfo {author} {\bibfnamefont {S.~V.}\ \bibnamefont {Morozov}}, \bibinfo
  {author} {\bibfnamefont {D.}~\bibnamefont {Jiang}}, \bibinfo {author}
  {\bibfnamefont {M.~I.}\ \bibnamefont {Katsnelson}}, \bibinfo {author}
  {\bibfnamefont {I.~V.}\ \bibnamefont {Grigorieva}}, \bibinfo {author}
  {\bibfnamefont {S.~V.}\ \bibnamefont {Dubonos}},\ and\ \bibinfo {author}
  {\bibfnamefont {A.~A.}\ \bibnamefont {Firsov}},\ }\href
  {https://doi.org/10.1038/nature04233} {\bibfield  {journal} {\bibinfo
  {journal} {Nature}\ }\textbf {\bibinfo {volume} {438}},\ \bibinfo {pages}
  {197} (\bibinfo {year} {2005})}\BibitemShut {NoStop}%
\bibitem [{\citenamefont {Castro~Neto}\ \emph {et~al.}(2009)\citenamefont
  {Castro~Neto}, \citenamefont {Guinea}, \citenamefont {Peres}, \citenamefont
  {Novoselov},\ and\ \citenamefont {Geim}}]{RMP.2009.81.109}%
  \BibitemOpen
  \bibfield  {author} {\bibinfo {author} {\bibfnamefont {A.~H.}\ \bibnamefont
  {Castro~Neto}}, \bibinfo {author} {\bibfnamefont {F.}~\bibnamefont {Guinea}},
  \bibinfo {author} {\bibfnamefont {N.~M.~R.}\ \bibnamefont {Peres}}, \bibinfo
  {author} {\bibfnamefont {K.~S.}\ \bibnamefont {Novoselov}},\ and\ \bibinfo
  {author} {\bibfnamefont {A.~K.}\ \bibnamefont {Geim}},\ }\href
  {https://doi.org/10.1103/RevModPhys.81.109} {\bibfield  {journal} {\bibinfo
  {journal} {Rev. Mod. Phys.}\ }\textbf {\bibinfo {volume} {81}},\ \bibinfo
  {pages} {109} (\bibinfo {year} {2009})}\BibitemShut {NoStop}%
\bibitem [{\citenamefont {Niu}\ \emph {et~al.}(2020)\citenamefont {Niu},
  \citenamefont {Wang}, \citenamefont {Zhang}, \citenamefont {Shen},
  \citenamefont {Wu}, \citenamefont {Zhang}, \citenamefont {Zhang},
  \citenamefont {Wang}, \citenamefont {Gao}, \citenamefont {Liu}, \citenamefont
  {Zhang}, \citenamefont {Yang}, \citenamefont {Wang},\ and\ \citenamefont
  {Xue}}]{Science.2020.367.864}%
  \BibitemOpen
  \bibfield  {author} {\bibinfo {author} {\bibfnamefont {D.}~\bibnamefont
  {Niu}}, \bibinfo {author} {\bibfnamefont {S.}~\bibnamefont {Wang}}, \bibinfo
  {author} {\bibfnamefont {G.}~\bibnamefont {Zhang}}, \bibinfo {author}
  {\bibfnamefont {X.}~\bibnamefont {Shen}}, \bibinfo {author} {\bibfnamefont
  {Z.}~\bibnamefont {Wu}}, \bibinfo {author} {\bibfnamefont {X.}~\bibnamefont
  {Zhang}}, \bibinfo {author} {\bibfnamefont {Z.}~\bibnamefont {Zhang}},
  \bibinfo {author} {\bibfnamefont {W.}~\bibnamefont {Wang}}, \bibinfo {author}
  {\bibfnamefont {P.}~\bibnamefont {Gao}}, \bibinfo {author} {\bibfnamefont
  {W.}~\bibnamefont {Liu}}, \bibinfo {author} {\bibfnamefont {C.}~\bibnamefont
  {Zhang}}, \bibinfo {author} {\bibfnamefont {J.}~\bibnamefont {Yang}},
  \bibinfo {author} {\bibfnamefont {H.}~\bibnamefont {Wang}},\ and\ \bibinfo
  {author} {\bibfnamefont {J.}~\bibnamefont {Xue}},\ }\href
  {https://doi.org/https://doi.org/10.1038/nphys2442} {\bibfield  {journal}
  {\bibinfo  {journal} {Science}\ }\textbf {\bibinfo {volume} {367}},\ \bibinfo
  {pages} {864} (\bibinfo {year} {2020})}\BibitemShut {NoStop}%
\bibitem [{\citenamefont {Tang}\ \emph {et~al.}(2019)\citenamefont {Tang},
  \citenamefont {Ren}, \citenamefont {Wang}, \citenamefont {Zhong},
  \citenamefont {Schneeloch}, \citenamefont {Yang}, \citenamefont {Yang},
  \citenamefont {Lee}, \citenamefont {Gu}, \citenamefont {Qiao},\ and\
  \citenamefont {Zhang}}]{Nature.2019.569.537}%
  \BibitemOpen
  \bibfield  {author} {\bibinfo {author} {\bibfnamefont {F.}~\bibnamefont
  {Tang}}, \bibinfo {author} {\bibfnamefont {Y.}~\bibnamefont {Ren}}, \bibinfo
  {author} {\bibfnamefont {P.}~\bibnamefont {Wang}}, \bibinfo {author}
  {\bibfnamefont {R.}~\bibnamefont {Zhong}}, \bibinfo {author} {\bibfnamefont
  {J.}~\bibnamefont {Schneeloch}}, \bibinfo {author} {\bibfnamefont {S.~A.}\
  \bibnamefont {Yang}}, \bibinfo {author} {\bibfnamefont {K.}~\bibnamefont
  {Yang}}, \bibinfo {author} {\bibfnamefont {P.~A.}\ \bibnamefont {Lee}},
  \bibinfo {author} {\bibfnamefont {G.}~\bibnamefont {Gu}}, \bibinfo {author}
  {\bibfnamefont {Z.}~\bibnamefont {Qiao}},\ and\ \bibinfo {author}
  {\bibfnamefont {L.}~\bibnamefont {Zhang}},\ }\href
  {https://doi.org/10.1038/s41586-019-1180-9} {\bibfield  {journal} {\bibinfo
  {journal} {Nature}\ }\textbf {\bibinfo {volume} {569}},\ \bibinfo {pages}
  {537} (\bibinfo {year} {2019})}\BibitemShut {NoStop}%
\bibitem [{\citenamefont {Shao}\ \emph {et~al.}(2024)\citenamefont {Shao},
  \citenamefont {Moon}, \citenamefont {Rudenko}, \citenamefont {Wang},
  \citenamefont {Herzog-Arbeitman}, \citenamefont {Ozerov}, \citenamefont
  {Graf}, \citenamefont {Sun}, \citenamefont {Queiroz}, \citenamefont {Lee},
  \citenamefont {Zhu}, \citenamefont {Mao}, \citenamefont {Katsnelson},
  \citenamefont {Bernevig}, \citenamefont {Smirnov}, \citenamefont {Millis},\
  and\ \citenamefont {Basov}}]{PhysRevX.14.041057}%
  \BibitemOpen
  \bibfield  {author} {\bibinfo {author} {\bibfnamefont {Y.}~\bibnamefont
  {Shao}}, \bibinfo {author} {\bibfnamefont {S.}~\bibnamefont {Moon}}, \bibinfo
  {author} {\bibfnamefont {A.~N.}\ \bibnamefont {Rudenko}}, \bibinfo {author}
  {\bibfnamefont {J.}~\bibnamefont {Wang}}, \bibinfo {author} {\bibfnamefont
  {J.}~\bibnamefont {Herzog-Arbeitman}}, \bibinfo {author} {\bibfnamefont
  {M.}~\bibnamefont {Ozerov}}, \bibinfo {author} {\bibfnamefont
  {D.}~\bibnamefont {Graf}}, \bibinfo {author} {\bibfnamefont {Z.}~\bibnamefont
  {Sun}}, \bibinfo {author} {\bibfnamefont {R.}~\bibnamefont {Queiroz}},
  \bibinfo {author} {\bibfnamefont {S.~H.}\ \bibnamefont {Lee}}, \bibinfo
  {author} {\bibfnamefont {Y.}~\bibnamefont {Zhu}}, \bibinfo {author}
  {\bibfnamefont {Z.}~\bibnamefont {Mao}}, \bibinfo {author} {\bibfnamefont
  {M.~I.}\ \bibnamefont {Katsnelson}}, \bibinfo {author} {\bibfnamefont
  {B.~A.}\ \bibnamefont {Bernevig}}, \bibinfo {author} {\bibfnamefont
  {D.}~\bibnamefont {Smirnov}}, \bibinfo {author} {\bibfnamefont {A.~J.}\
  \bibnamefont {Millis}},\ and\ \bibinfo {author} {\bibfnamefont {D.~N.}\
  \bibnamefont {Basov}},\ }\href {https://doi.org/10.1103/PhysRevX.14.041057}
  {\bibfield  {journal} {\bibinfo  {journal} {Phys. Rev. X}\ }\textbf {\bibinfo
  {volume} {14}},\ \bibinfo {pages} {041057} (\bibinfo {year}
  {2024})}\BibitemShut {NoStop}%
\bibitem [{\citenamefont {Di~Francesco}\ \emph {et~al.}(1997)\citenamefont
  {Di~Francesco}, \citenamefont {Mathieu},\ and\ \citenamefont
  {Sénéchal}}]{DiFrancesco1997}%
  \BibitemOpen
  \bibfield  {author} {\bibinfo {author} {\bibfnamefont {P.}~\bibnamefont
  {Di~Francesco}}, \bibinfo {author} {\bibfnamefont {P.}~\bibnamefont
  {Mathieu}},\ and\ \bibinfo {author} {\bibfnamefont {D.}~\bibnamefont
  {Sénéchal}},\ }\href@noop {} {\emph {\bibinfo {title} {Conformal Field
  Theory}}}\ (\bibinfo  {publisher} {Springer},\ \bibinfo {address} {Science \&
  Business Media},\ \bibinfo {year} {1997})\BibitemShut {NoStop}%
\bibitem [{\citenamefont {Maldacena}(1999)}]{IJTP.1999.38.1113}%
  \BibitemOpen
  \bibfield  {author} {\bibinfo {author} {\bibfnamefont {J.}~\bibnamefont
  {Maldacena}},\ }\href {https://doi.org/10.1023/A:1026654312961} {\bibfield
  {journal} {\bibinfo  {journal} {International Journal of Theoretical
  Physics}\ }\textbf {\bibinfo {volume} {38}},\ \bibinfo {pages} {1113}
  (\bibinfo {year} {1999})}\BibitemShut {NoStop}%
\bibitem [{\citenamefont {Gubser}\ \emph {et~al.}(1998)\citenamefont {Gubser},
  \citenamefont {Klebanov},\ and\ \citenamefont {Polyakov}}]{PLB.1998.428.105}%
  \BibitemOpen
  \bibfield  {author} {\bibinfo {author} {\bibfnamefont {S.}~\bibnamefont
  {Gubser}}, \bibinfo {author} {\bibfnamefont {I.}~\bibnamefont {Klebanov}},\
  and\ \bibinfo {author} {\bibfnamefont {A.}~\bibnamefont {Polyakov}},\ }\href
  {https://doi.org/https://doi.org/10.1016/S0370-2693(98)00377-3} {\bibfield
  {journal} {\bibinfo  {journal} {Physics Letters B}\ }\textbf {\bibinfo
  {volume} {428}},\ \bibinfo {pages} {105} (\bibinfo {year}
  {1998})}\BibitemShut {NoStop}%
\bibitem [{\citenamefont {Witten}(1998)}]{ATMP.1998.2.253}%
  \BibitemOpen
  \bibfield  {author} {\bibinfo {author} {\bibfnamefont {E.}~\bibnamefont
  {Witten}},\ }\href {https://doi.org/10.4310/atmp.1998.v2.n2.a2} {\bibfield
  {journal} {\bibinfo  {journal} {Advances in Theoretical and Mathematical
  Physics}\ }\textbf {\bibinfo {volume} {2}},\ \bibinfo {pages} {253} (\bibinfo
  {year} {1998})}\BibitemShut {NoStop}%
\bibitem [{\citenamefont {Kosteleck\'y}\ and\ \citenamefont
  {Samuel}(1989)}]{PRD.1989.39.683}%
  \BibitemOpen
  \bibfield  {author} {\bibinfo {author} {\bibfnamefont {V.~A.}\ \bibnamefont
  {Kosteleck\'y}}\ and\ \bibinfo {author} {\bibfnamefont {S.}~\bibnamefont
  {Samuel}},\ }\href {https://doi.org/10.1103/PhysRevD.39.683} {\bibfield
  {journal} {\bibinfo  {journal} {Phys. Rev. D}\ }\textbf {\bibinfo {volume}
  {39}},\ \bibinfo {pages} {683} (\bibinfo {year} {1989})}\BibitemShut
  {NoStop}%
\bibitem [{\citenamefont {Jackiw}\ and\ \citenamefont
  {Kosteleck\'y}(1999)}]{PRL.1999.82.3572}%
  \BibitemOpen
  \bibfield  {author} {\bibinfo {author} {\bibfnamefont {R.}~\bibnamefont
  {Jackiw}}\ and\ \bibinfo {author} {\bibfnamefont {V.~A.}\ \bibnamefont
  {Kosteleck\'y}},\ }\href {https://doi.org/10.1103/PhysRevLett.82.3572}
  {\bibfield  {journal} {\bibinfo  {journal} {Phys. Rev. Lett.}\ }\textbf
  {\bibinfo {volume} {82}},\ \bibinfo {pages} {3572} (\bibinfo {year}
  {1999})}\BibitemShut {NoStop}%
\bibitem [{\citenamefont {Colladay}\ and\ \citenamefont
  {Kosteleck\'y}(1997)}]{PRD.1997.55.6760}%
  \BibitemOpen
  \bibfield  {author} {\bibinfo {author} {\bibfnamefont {D.}~\bibnamefont
  {Colladay}}\ and\ \bibinfo {author} {\bibfnamefont {V.~A.}\ \bibnamefont
  {Kosteleck\'y}},\ }\href {https://doi.org/10.1103/PhysRevD.55.6760}
  {\bibfield  {journal} {\bibinfo  {journal} {Phys. Rev. D}\ }\textbf {\bibinfo
  {volume} {55}},\ \bibinfo {pages} {6760} (\bibinfo {year}
  {1997})}\BibitemShut {NoStop}%
\bibitem [{\citenamefont {Kostelecký}\ and\ \citenamefont
  {Lane}(1999)}]{JMP.1999.40.6245}%
  \BibitemOpen
  \bibfield  {author} {\bibinfo {author} {\bibfnamefont {V.~A.}\ \bibnamefont
  {Kostelecký}}\ and\ \bibinfo {author} {\bibfnamefont {C.~D.}\ \bibnamefont
  {Lane}},\ }\href {https://doi.org/10.1063/1.533090} {\bibfield  {journal}
  {\bibinfo  {journal} {Journal of Mathematical Physics}\ }\textbf {\bibinfo
  {volume} {40}},\ \bibinfo {pages} {6245} (\bibinfo {year}
  {1999})}\BibitemShut {NoStop}%
\bibitem [{\citenamefont {Kostelecký}\ and\ \citenamefont
  {Potting}(1991)}]{NPB.1991.359.545}%
  \BibitemOpen
  \bibfield  {author} {\bibinfo {author} {\bibfnamefont {V.}~\bibnamefont
  {Kostelecký}}\ and\ \bibinfo {author} {\bibfnamefont {R.}~\bibnamefont
  {Potting}},\ }\href {https://doi.org/10.1016/0550-3213(91)90071-5} {\bibfield
   {journal} {\bibinfo  {journal} {Nucl. Phys. B}\ }\textbf {\bibinfo {volume}
  {359}},\ \bibinfo {pages} {545} (\bibinfo {year} {1991})}\BibitemShut
  {NoStop}%
\bibitem [{\citenamefont {Bear}\ \emph {et~al.}(2000)\citenamefont {Bear},
  \citenamefont {Stoner}, \citenamefont {Walsworth}, \citenamefont
  {Kostelecký},\ and\ \citenamefont {Lane}}]{PRL.2000.85.5038}%
  \BibitemOpen
  \bibfield  {author} {\bibinfo {author} {\bibfnamefont {D.}~\bibnamefont
  {Bear}}, \bibinfo {author} {\bibfnamefont {R.}~\bibnamefont {Stoner}},
  \bibinfo {author} {\bibfnamefont {R.}~\bibnamefont {Walsworth}}, \bibinfo
  {author} {\bibfnamefont {V.}~\bibnamefont {Kostelecký}},\ and\ \bibinfo
  {author} {\bibfnamefont {C.}~\bibnamefont {Lane}},\ }\href
  {https://doi.org/10.1103/PhysRevLett.85.5038} {\bibfield  {journal} {\bibinfo
   {journal} {Phys. Rev. Lett.}\ }\textbf {\bibinfo {volume} {85}},\ \bibinfo
  {pages} {5038} (\bibinfo {year} {2000})}\BibitemShut {NoStop}%
\bibitem [{\citenamefont {Kostelecký}\ and\ \citenamefont
  {Mewes}(2002)}]{PRD.2002.66.056005}%
  \BibitemOpen
  \bibfield  {author} {\bibinfo {author} {\bibfnamefont {V.}~\bibnamefont
  {Kostelecký}}\ and\ \bibinfo {author} {\bibfnamefont {M.}~\bibnamefont
  {Mewes}},\ }\href {https://doi.org/10.1103/PhysRevD.66.056005} {\bibfield
  {journal} {\bibinfo  {journal} {Phys. Rev. D}\ }\textbf {\bibinfo {volume}
  {66}},\ \bibinfo {pages} {056005} (\bibinfo {year} {2002})}\BibitemShut
  {NoStop}%
\bibitem [{\citenamefont {Humphrey}\ \emph {et~al.}(2000)\citenamefont
  {Humphrey}, \citenamefont {Phillips}, \citenamefont {Mattison}, \citenamefont
  {Vessot}, \citenamefont {Walsworth}, \citenamefont {Kostelecký},\ and\
  \citenamefont {Lane}}]{PRA.2000.62.063405}%
  \BibitemOpen
  \bibfield  {author} {\bibinfo {author} {\bibfnamefont {M.}~\bibnamefont
  {Humphrey}}, \bibinfo {author} {\bibfnamefont {D.}~\bibnamefont {Phillips}},
  \bibinfo {author} {\bibfnamefont {E.}~\bibnamefont {Mattison}}, \bibinfo
  {author} {\bibfnamefont {R.}~\bibnamefont {Vessot}}, \bibinfo {author}
  {\bibfnamefont {R.}~\bibnamefont {Walsworth}}, \bibinfo {author}
  {\bibfnamefont {V.}~\bibnamefont {Kostelecký}},\ and\ \bibinfo {author}
  {\bibfnamefont {C.}~\bibnamefont {Lane}},\ }\href
  {https://doi.org/10.1103/PhysRevA.62.063405} {\bibfield  {journal} {\bibinfo
  {journal} {Phys. Rev. A}\ }\textbf {\bibinfo {volume} {62}},\ \bibinfo
  {pages} {063405} (\bibinfo {year} {2000})}\BibitemShut {NoStop}%
\bibitem [{\citenamefont {Lane}(2006)}]{PRD.2006.72.016005}%
  \BibitemOpen
  \bibfield  {author} {\bibinfo {author} {\bibfnamefont {C.}~\bibnamefont
  {Lane}},\ }\href {https://doi.org/10.1103/PhysRevD.72.016005} {\bibfield
  {journal} {\bibinfo  {journal} {Phys. Rev. D}\ }\textbf {\bibinfo {volume}
  {72}},\ \bibinfo {pages} {016005} (\bibinfo {year} {2006})}\BibitemShut
  {NoStop}%
\bibitem [{\citenamefont {Mewes}(2011)}]{Kostelecky2011}%
  \BibitemOpen
  \bibfield  {author} {\bibinfo {author} {\bibfnamefont {M.}~\bibnamefont
  {Mewes}},\ }\href
  {https://doi.org/https://doi.org/10.1016/j.nuclphysbps.2011.10.021}
  {\bibfield  {journal} {\bibinfo  {journal} {Nuclear Physics B - Proceedings
  Supplements}\ }\textbf {\bibinfo {volume} {221}},\ \bibinfo {pages} {373}
  (\bibinfo {year} {2011})}\BibitemShut {NoStop}%
\bibitem [{\citenamefont {D\'{\i}az}\ \emph {et~al.}(2014)\citenamefont
  {D\'{\i}az}, \citenamefont {Kosteleck\'y},\ and\ \citenamefont
  {Mewes}}]{Kostelecky2016}%
  \BibitemOpen
  \bibfield  {author} {\bibinfo {author} {\bibfnamefont {J.~S.}\ \bibnamefont
  {D\'{\i}az}}, \bibinfo {author} {\bibfnamefont {V.~A.}\ \bibnamefont
  {Kosteleck\'y}},\ and\ \bibinfo {author} {\bibfnamefont {M.}~\bibnamefont
  {Mewes}},\ }\href {https://doi.org/10.1103/PhysRevD.89.043005} {\bibfield
  {journal} {\bibinfo  {journal} {Phys. Rev. D}\ }\textbf {\bibinfo {volume}
  {89}},\ \bibinfo {pages} {043005} (\bibinfo {year} {2014})}\BibitemShut
  {NoStop}%
\bibitem [{\citenamefont {Kostelecký}\ and\ \citenamefont
  {Tasson}(2015)}]{Kostelecky2021}%
  \BibitemOpen
  \bibfield  {author} {\bibinfo {author} {\bibfnamefont {V.~A.}\ \bibnamefont
  {Kostelecký}}\ and\ \bibinfo {author} {\bibfnamefont {J.~D.}\ \bibnamefont
  {Tasson}},\ }\href
  {https://doi.org/https://doi.org/10.1016/j.physletb.2015.08.060} {\bibfield
  {journal} {\bibinfo  {journal} {Physics Letters B}\ }\textbf {\bibinfo
  {volume} {749}},\ \bibinfo {pages} {551} (\bibinfo {year}
  {2015})}\BibitemShut {NoStop}%
\bibitem [{\citenamefont {Safronova}\ \emph {et~al.}(2018)\citenamefont
  {Safronova}, \citenamefont {Budker}, \citenamefont {DeMille}, \citenamefont
  {Kimball}, \citenamefont {Derevianko},\ and\ \citenamefont
  {Clark}}]{Safronova2018}%
  \BibitemOpen
  \bibfield  {author} {\bibinfo {author} {\bibfnamefont {M.~S.}\ \bibnamefont
  {Safronova}}, \bibinfo {author} {\bibfnamefont {D.}~\bibnamefont {Budker}},
  \bibinfo {author} {\bibfnamefont {D.}~\bibnamefont {DeMille}}, \bibinfo
  {author} {\bibfnamefont {D.~F.~J.}\ \bibnamefont {Kimball}}, \bibinfo
  {author} {\bibfnamefont {A.}~\bibnamefont {Derevianko}},\ and\ \bibinfo
  {author} {\bibfnamefont {C.~W.}\ \bibnamefont {Clark}},\ }\href
  {https://doi.org/10.1103/RevModPhys.90.025008} {\bibfield  {journal}
  {\bibinfo  {journal} {Rev. Mod. Phys.}\ }\textbf {\bibinfo {volume} {90}},\
  \bibinfo {pages} {025008} (\bibinfo {year} {2018})}\BibitemShut {NoStop}%
\bibitem [{\citenamefont {Borges}\ \emph {et~al.}(2024)\citenamefont {Borges},
  \citenamefont {Ferrari}, \citenamefont {{da Silva}},\ and\ \citenamefont
  {Barone}}]{borges2024}%
  \BibitemOpen
  \bibfield  {author} {\bibinfo {author} {\bibfnamefont {L.}~\bibnamefont
  {Borges}}, \bibinfo {author} {\bibfnamefont {A.}~\bibnamefont {Ferrari}},
  \bibinfo {author} {\bibfnamefont {P.}~\bibnamefont {{da Silva}}},\ and\
  \bibinfo {author} {\bibfnamefont {F.}~\bibnamefont {Barone}},\ }\href
  {https://doi.org/https://doi.org/10.1016/j.aop.2024.169749} {\bibfield
  {journal} {\bibinfo  {journal} {Annals of Physics}\ }\textbf {\bibinfo
  {volume} {469}},\ \bibinfo {pages} {169749} (\bibinfo {year}
  {2024})}\BibitemShut {NoStop}%
\bibitem [{\citenamefont {Jackiw}\ and\ \citenamefont
  {Pi}(2003)}]{jackiw2003chern}%
  \BibitemOpen
  \bibfield  {author} {\bibinfo {author} {\bibfnamefont {R.}~\bibnamefont
  {Jackiw}}\ and\ \bibinfo {author} {\bibfnamefont {S.-Y.}\ \bibnamefont
  {Pi}},\ }\href {https://doi.org/10.1103/PhysRevD.68.104012} {\bibfield
  {journal} {\bibinfo  {journal} {Physical Review D}\ }\textbf {\bibinfo
  {volume} {68}},\ \bibinfo {pages} {104012} (\bibinfo {year}
  {2003})}\BibitemShut {NoStop}%
\bibitem [{\citenamefont {Nair}\ and\ \citenamefont
  {Polychronakos}(2001)}]{nair2001massive}%
  \BibitemOpen
  \bibfield  {author} {\bibinfo {author} {\bibfnamefont {V.~P.}\ \bibnamefont
  {Nair}}\ and\ \bibinfo {author} {\bibfnamefont {A.~P.}\ \bibnamefont
  {Polychronakos}},\ }\href {https://doi.org/10.1103/PhysRevLett.87.030403}
  {\bibfield  {journal} {\bibinfo  {journal} {Physical Review Letters}\
  }\textbf {\bibinfo {volume} {87}},\ \bibinfo {pages} {030403} (\bibinfo
  {year} {2001})}\BibitemShut {NoStop}%
\bibitem [{\citenamefont {Deser}\ \emph {et~al.}(1982)\citenamefont {Deser},
  \citenamefont {Jackiw},\ and\ \citenamefont {Templeton}}]{DESER1982372}%
  \BibitemOpen
  \bibfield  {author} {\bibinfo {author} {\bibfnamefont {S.}~\bibnamefont
  {Deser}}, \bibinfo {author} {\bibfnamefont {R.}~\bibnamefont {Jackiw}},\ and\
  \bibinfo {author} {\bibfnamefont {S.}~\bibnamefont {Templeton}},\ }\href
  {https://doi.org/https://doi.org/10.1016/0003-4916(82)90164-6} {\bibfield
  {journal} {\bibinfo  {journal} {Annals of Physics}\ }\textbf {\bibinfo
  {volume} {140}},\ \bibinfo {pages} {372} (\bibinfo {year}
  {1982})}\BibitemShut {NoStop}%
\bibitem [{\citenamefont {Reis}\ \emph {et~al.}(2019)\citenamefont {Reis},
  \citenamefont {Ferreira},\ and\ \citenamefont {Schreck}}]{schreck2019}%
  \BibitemOpen
  \bibfield  {author} {\bibinfo {author} {\bibfnamefont {J.~A. A.~S.}\
  \bibnamefont {Reis}}, \bibinfo {author} {\bibfnamefont {M.~M.}\ \bibnamefont
  {Ferreira}},\ and\ \bibinfo {author} {\bibfnamefont {M.}~\bibnamefont
  {Schreck}},\ }\href {https://doi.org/10.1103/PhysRevD.100.095026} {\bibfield
  {journal} {\bibinfo  {journal} {Phys. Rev. D}\ }\textbf {\bibinfo {volume}
  {100}},\ \bibinfo {pages} {095026} (\bibinfo {year} {2019})}\BibitemShut
  {NoStop}%
\bibitem [{\citenamefont {Lisboa-Santos}\ \emph
  {et~al.}(2023{\natexlab{a}})\citenamefont {Lisboa-Santos}, \citenamefont
  {Reis}, \citenamefont {Schreck},\ and\ \citenamefont
  {Ferreira}}]{schreck2023}%
  \BibitemOpen
  \bibfield  {author} {\bibinfo {author} {\bibfnamefont {L.}~\bibnamefont
  {Lisboa-Santos}}, \bibinfo {author} {\bibfnamefont {J.~A. A.~S.}\
  \bibnamefont {Reis}}, \bibinfo {author} {\bibfnamefont {M.}~\bibnamefont
  {Schreck}},\ and\ \bibinfo {author} {\bibfnamefont {M.~M.}\ \bibnamefont
  {Ferreira}},\ }\href {https://doi.org/10.1103/PhysRevD.108.115032} {\bibfield
   {journal} {\bibinfo  {journal} {Phys. Rev. D}\ }\textbf {\bibinfo {volume}
  {108}},\ \bibinfo {pages} {115032} (\bibinfo {year}
  {2023}{\natexlab{a}})}\BibitemShut {NoStop}%
\bibitem [{\citenamefont {Ping{ }He}\ and\ \citenamefont {{Bo-Qiang
  Ma}}(2022)}]{Tasson2022}%
  \BibitemOpen
  \bibfield  {author} {\bibinfo {author} {\bibnamefont {Ping{ }He}}\ and\
  \bibinfo {author} {\bibnamefont {{Bo-Qiang Ma}}},\ }\href
  {https://doi.org/10.3390/universe8060323} {\bibfield  {journal} {\bibinfo
  {journal} {Universe}\ }\textbf {\bibinfo {volume} {8}},\ \bibinfo {pages}
  {323} (\bibinfo {year} {2022})}\BibitemShut {NoStop}%
\bibitem [{\citenamefont {D\'{\i}az}\ \emph {et~al.}(2013)\citenamefont
  {D\'{\i}az}, \citenamefont {Kosteleck\'y},\ and\ \citenamefont
  {Lehnert}}]{Diaz2023}%
  \BibitemOpen
  \bibfield  {author} {\bibinfo {author} {\bibfnamefont {J.~S.}\ \bibnamefont
  {D\'{\i}az}}, \bibinfo {author} {\bibfnamefont {V.~A.}\ \bibnamefont
  {Kosteleck\'y}},\ and\ \bibinfo {author} {\bibfnamefont {R.}~\bibnamefont
  {Lehnert}},\ }\href {https://doi.org/10.1103/PhysRevD.88.071902} {\bibfield
  {journal} {\bibinfo  {journal} {Phys. Rev. D}\ }\textbf {\bibinfo {volume}
  {88}},\ \bibinfo {pages} {071902} (\bibinfo {year} {2013})}\BibitemShut
  {NoStop}%
\bibitem [{\citenamefont {Alan~Kosteleck\'y}\ and\ \citenamefont
  {Mewes}(2004)}]{Kostelecky2024}%
  \BibitemOpen
  \bibfield  {author} {\bibinfo {author} {\bibfnamefont {V.}~\bibnamefont
  {Alan~Kosteleck\'y}}\ and\ \bibinfo {author} {\bibfnamefont {M.}~\bibnamefont
  {Mewes}},\ }\href {https://doi.org/10.1103/PhysRevD.69.016005} {\bibfield
  {journal} {\bibinfo  {journal} {Phys. Rev. D}\ }\textbf {\bibinfo {volume}
  {69}},\ \bibinfo {pages} {016005} (\bibinfo {year} {2004})}\BibitemShut
  {NoStop}%
\bibitem [{\citenamefont {Kosteleck\'y}\ and\ \citenamefont
  {Mewes}(2002)}]{Kostelecky2002}%
  \BibitemOpen
  \bibfield  {author} {\bibinfo {author} {\bibfnamefont {V.~A.}\ \bibnamefont
  {Kosteleck\'y}}\ and\ \bibinfo {author} {\bibfnamefont {M.}~\bibnamefont
  {Mewes}},\ }\href {https://doi.org/10.1103/PhysRevD.66.056005} {\bibfield
  {journal} {\bibinfo  {journal} {Phys. Rev. D}\ }\textbf {\bibinfo {volume}
  {66}},\ \bibinfo {pages} {056005} (\bibinfo {year} {2002})}\BibitemShut
  {NoStop}%
\bibitem [{\citenamefont {Altschul}(2006)}]{Altschul2006}%
  \BibitemOpen
  \bibfield  {author} {\bibinfo {author} {\bibfnamefont {B.}~\bibnamefont
  {Altschul}},\ }\href {https://doi.org/10.1103/PhysRevD.73.036005} {\bibfield
  {journal} {\bibinfo  {journal} {Phys. Rev. D}\ }\textbf {\bibinfo {volume}
  {73}},\ \bibinfo {pages} {036005} (\bibinfo {year} {2006})}\BibitemShut
  {NoStop}%
\bibitem [{\citenamefont {Carroll}\ \emph {et~al.}(2001)\citenamefont
  {Carroll}, \citenamefont {Harvey}, \citenamefont {Kostelecký}, \citenamefont
  {Lane},\ and\ \citenamefont {Okamoto}}]{Carroll2001}%
  \BibitemOpen
  \bibfield  {author} {\bibinfo {author} {\bibfnamefont {S.}~\bibnamefont
  {Carroll}}, \bibinfo {author} {\bibfnamefont {J.}~\bibnamefont {Harvey}},
  \bibinfo {author} {\bibfnamefont {V.}~\bibnamefont {Kostelecký}}, \bibinfo
  {author} {\bibfnamefont {C.}~\bibnamefont {Lane}},\ and\ \bibinfo {author}
  {\bibfnamefont {T.}~\bibnamefont {Okamoto}},\ }\href
  {https://doi.org/https://doi.org/10.1103/PhysRevLett.87.141601} {\bibfield
  {journal} {\bibinfo  {journal} {Phys. Rev. Lett.}\ }\textbf {\bibinfo
  {volume} {87}},\ \bibinfo {pages} {141601} (\bibinfo {year}
  {2001})}\BibitemShut {NoStop}%
\bibitem [{\citenamefont {Casana}\ \emph
  {et~al.}(2013{\natexlab{a}})\citenamefont {Casana}, \citenamefont
  {Ferreira~Jr.}, \citenamefont {Maluf},\ and\ \citenamefont
  {Dos~Santos}}]{casana2013radiative}%
  \BibitemOpen
  \bibfield  {author} {\bibinfo {author} {\bibfnamefont {R.}~\bibnamefont
  {Casana}}, \bibinfo {author} {\bibfnamefont {M.~M.}\ \bibnamefont
  {Ferreira~Jr.}}, \bibinfo {author} {\bibfnamefont {R.~V.}\ \bibnamefont
  {Maluf}},\ and\ \bibinfo {author} {\bibfnamefont {F.~E.~P.}\ \bibnamefont
  {Dos~Santos}},\ }\href {https://doi.org/10.1016/j.physletb.2013.04.003}
  {\bibfield  {journal} {\bibinfo  {journal} {Physics Letters B}\ }\textbf
  {\bibinfo {volume} {722}},\ \bibinfo {pages} {315} (\bibinfo {year}
  {2013}{\natexlab{a}})}\BibitemShut {NoStop}%
\bibitem [{\citenamefont {Araujo}\ \emph {et~al.}(2016)\citenamefont {Araujo},
  \citenamefont {Casana},\ and\ \citenamefont
  {Ferreira~Jr.}}]{araujo2016general}%
  \BibitemOpen
  \bibfield  {author} {\bibinfo {author} {\bibfnamefont {J.~B.}\ \bibnamefont
  {Araujo}}, \bibinfo {author} {\bibfnamefont {R.}~\bibnamefont {Casana}},\
  and\ \bibinfo {author} {\bibfnamefont {M.~M.}\ \bibnamefont {Ferreira~Jr.}},\
  }\href {https://doi.org/10.1016/j.physletb.2016.07.004} {\bibfield  {journal}
  {\bibinfo  {journal} {Physics Letters B}\ }\textbf {\bibinfo {volume}
  {760}},\ \bibinfo {pages} {302} (\bibinfo {year} {2016})}\BibitemShut
  {NoStop}%
\bibitem [{\citenamefont {Casana}\ \emph
  {et~al.}(2013{\natexlab{b}})\citenamefont {Casana}, \citenamefont
  {Ferreira~Jr.},\ and\ \citenamefont {Passos}}]{casana2013new}%
  \BibitemOpen
  \bibfield  {author} {\bibinfo {author} {\bibfnamefont {R.}~\bibnamefont
  {Casana}}, \bibinfo {author} {\bibfnamefont {M.~M.}\ \bibnamefont
  {Ferreira~Jr.}},\ and\ \bibinfo {author} {\bibfnamefont {E.}~\bibnamefont
  {Passos}},\ }\href {https://doi.org/10.1103/PhysRevD.87.047701} {\bibfield
  {journal} {\bibinfo  {journal} {Physical Review D}\ }\textbf {\bibinfo
  {volume} {87}},\ \bibinfo {pages} {047701} (\bibinfo {year}
  {2013}{\natexlab{b}})}\BibitemShut {NoStop}%
\bibitem [{\citenamefont {Carvalho}\ \emph {et~al.}(2023)\citenamefont
  {Carvalho}, \citenamefont {Dias},\ and\ \citenamefont
  {Lehum}}]{carvalho2023perturbative}%
  \BibitemOpen
  \bibfield  {author} {\bibinfo {author} {\bibfnamefont {W.}~\bibnamefont
  {Carvalho}}, \bibinfo {author} {\bibfnamefont {M.}~\bibnamefont {Dias}},\
  and\ \bibinfo {author} {\bibfnamefont {A.~C.}\ \bibnamefont {Lehum}},\ }\href
  {https://arxiv.org/pdf/2305.16848} {\bibfield  {journal} {\bibinfo  {journal}
  {Europhysics Letters}\ } (\bibinfo {year} {2023})}\BibitemShut {NoStop}%
\bibitem [{\citenamefont {Gazzola}\ \emph {et~al.}(2012)\citenamefont
  {Gazzola}, \citenamefont {Fargnoli},\ and\ \citenamefont
  {Scarpelli}}]{gazzola2012qed}%
  \BibitemOpen
  \bibfield  {author} {\bibinfo {author} {\bibfnamefont {G.}~\bibnamefont
  {Gazzola}}, \bibinfo {author} {\bibfnamefont {H.~G.}\ \bibnamefont
  {Fargnoli}},\ and\ \bibinfo {author} {\bibfnamefont {A.~P.~B.}\ \bibnamefont
  {Scarpelli}},\ }\href {https://doi.org/10.1088/0954-3899/39/3/035002}
  {\bibfield  {journal} {\bibinfo  {journal} {Journal of Physics G}\ }\textbf
  {\bibinfo {volume} {39}},\ \bibinfo {pages} {035002} (\bibinfo {year}
  {2012})}\BibitemShut {NoStop}%
\bibitem [{\citenamefont {Fabbri}\ \emph {et~al.}(2014)\citenamefont {Fabbri},
  \citenamefont {Vignolo},\ and\ \citenamefont
  {Carloni}}]{fabbri2014renormalizability}%
  \BibitemOpen
  \bibfield  {author} {\bibinfo {author} {\bibfnamefont {L.}~\bibnamefont
  {Fabbri}}, \bibinfo {author} {\bibfnamefont {S.}~\bibnamefont {Vignolo}},\
  and\ \bibinfo {author} {\bibfnamefont {S.}~\bibnamefont {Carloni}},\ }\href
  {https://doi.org/10.1103/PhysRevD.90.024012} {\bibfield  {journal} {\bibinfo
  {journal} {Physical Review D}\ }\textbf {\bibinfo {volume} {90}},\ \bibinfo
  {pages} {024012} (\bibinfo {year} {2014})}\BibitemShut {NoStop}%
\bibitem [{\citenamefont {Griguolo}\ and\ \citenamefont
  {Seminara}(1997)}]{griguolo1997nonminimal}%
  \BibitemOpen
  \bibfield  {author} {\bibinfo {author} {\bibfnamefont {L.}~\bibnamefont
  {Griguolo}}\ and\ \bibinfo {author} {\bibfnamefont {D.}~\bibnamefont
  {Seminara}},\ }\href {https://doi.org/10.1016/S0550-3213(97)00209-5}
  {\bibfield  {journal} {\bibinfo  {journal} {Nuclear Physics B}\ }\textbf
  {\bibinfo {volume} {504}},\ \bibinfo {pages} {615} (\bibinfo {year}
  {1997})}\BibitemShut {NoStop}%
\bibitem [{\citenamefont {Alonso}\ \emph {et~al.}(2014)\citenamefont {Alonso},
  \citenamefont {Jenkins}, \citenamefont {Manohar},\ and\ \citenamefont
  {Trott}}]{alonso2014renormalization}%
  \BibitemOpen
  \bibfield  {author} {\bibinfo {author} {\bibfnamefont {R.}~\bibnamefont
  {Alonso}}, \bibinfo {author} {\bibfnamefont {E.~E.}\ \bibnamefont {Jenkins}},
  \bibinfo {author} {\bibfnamefont {A.~V.}\ \bibnamefont {Manohar}},\ and\
  \bibinfo {author} {\bibfnamefont {M.}~\bibnamefont {Trott}},\ }\href
  {https://doi.org/10.1007/JHEP04(2014)159} {\bibfield  {journal} {\bibinfo
  {journal} {Journal of High Energy Physics}\ }\textbf {\bibinfo {volume}
  {2014}},\ \bibinfo {pages} {159} (\bibinfo {year} {2014})}\BibitemShut
  {NoStop}%
\bibitem [{\citenamefont {Kosteleck\'y}\ \emph {et~al.}(2022)\citenamefont
  {Kosteleck\'y}, \citenamefont {Lehnert}, \citenamefont {McGinnis},
  \citenamefont {Schreck},\ and\ \citenamefont
  {Seradjeh}}]{PhysRevResearch.Schreck}%
  \BibitemOpen
  \bibfield  {author} {\bibinfo {author} {\bibfnamefont {V.~A.}\ \bibnamefont
  {Kosteleck\'y}}, \bibinfo {author} {\bibfnamefont {R.}~\bibnamefont
  {Lehnert}}, \bibinfo {author} {\bibfnamefont {N.}~\bibnamefont {McGinnis}},
  \bibinfo {author} {\bibfnamefont {M.}~\bibnamefont {Schreck}},\ and\ \bibinfo
  {author} {\bibfnamefont {B.}~\bibnamefont {Seradjeh}},\ }\href
  {https://doi.org/10.1103/PhysRevResearch.4.023106} {\bibfield  {journal}
  {\bibinfo  {journal} {Phys. Rev. Res.}\ }\textbf {\bibinfo {volume} {4}},\
  \bibinfo {pages} {023106} (\bibinfo {year} {2022})}\BibitemShut {NoStop}%
\bibitem [{\citenamefont {Eichhorn}\ and\ \citenamefont
  {Lippoldt}(2017)}]{eichhorn2017quantum}%
  \BibitemOpen
  \bibfield  {author} {\bibinfo {author} {\bibfnamefont {A.}~\bibnamefont
  {Eichhorn}}\ and\ \bibinfo {author} {\bibfnamefont {S.}~\bibnamefont
  {Lippoldt}},\ }\href {https://doi.org/10.1016/j.physletb.2017.01.047}
  {\bibfield  {journal} {\bibinfo  {journal} {Physics Letters B}\ }\textbf
  {\bibinfo {volume} {767}},\ \bibinfo {pages} {142} (\bibinfo {year}
  {2017})}\BibitemShut {NoStop}%
\bibitem [{\citenamefont {Bettoni}\ and\ \citenamefont
  {Rubio}(2018)}]{bettoni2018quintessential}%
  \BibitemOpen
  \bibfield  {author} {\bibinfo {author} {\bibfnamefont {D.}~\bibnamefont
  {Bettoni}}\ and\ \bibinfo {author} {\bibfnamefont {J.}~\bibnamefont
  {Rubio}},\ }\href {https://doi.org/10.1016/j.physletb.2018.07.064} {\bibfield
   {journal} {\bibinfo  {journal} {Physics Letters B}\ }\textbf {\bibinfo
  {volume} {784}},\ \bibinfo {pages} {176} (\bibinfo {year}
  {2018})}\BibitemShut {NoStop}%
\bibitem [{\citenamefont {Wahlang}(2022)}]{wahlang2022evolution}%
  \BibitemOpen
  \bibfield  {author} {\bibinfo {author} {\bibfnamefont {W.}~\bibnamefont
  {Wahlang}},\ }\href {https://doi.org/10.1140/epjc/s10052-022-10304-1}
  {\bibfield  {journal} {\bibinfo  {journal} {The European Physical Journal C}\
  }\textbf {\bibinfo {volume} {82}},\ \bibinfo {pages} {305} (\bibinfo {year}
  {2022})}\BibitemShut {NoStop}%
\bibitem [{\citenamefont {Belfiglio}\ \emph {et~al.}(2024)\citenamefont
  {Belfiglio}, \citenamefont {Carloni},\ and\ \citenamefont
  {Luongo}}]{carloni2024particle}%
  \BibitemOpen
  \bibfield  {author} {\bibinfo {author} {\bibfnamefont {A.}~\bibnamefont
  {Belfiglio}}, \bibinfo {author} {\bibfnamefont {Y.}~\bibnamefont {Carloni}},\
  and\ \bibinfo {author} {\bibfnamefont {O.}~\bibnamefont {Luongo}},\ }\href
  {https://arxiv.org/pdf/2307.04739} {\bibfield  {journal} {\bibinfo  {journal}
  {Physics of the Dark Universe}\ } (\bibinfo {year} {2024})}\BibitemShut
  {NoStop}%
\bibitem [{\citenamefont {Rattazzi}\ and\ \citenamefont
  {Sarid}(1996)}]{rattazzi1996unified}%
  \BibitemOpen
  \bibfield  {author} {\bibinfo {author} {\bibfnamefont {R.}~\bibnamefont
  {Rattazzi}}\ and\ \bibinfo {author} {\bibfnamefont {U.}~\bibnamefont
  {Sarid}},\ }\href {https://doi.org/10.1103/PhysRevD.53.1553} {\bibfield
  {journal} {\bibinfo  {journal} {Physical Review D}\ }\textbf {\bibinfo
  {volume} {53}},\ \bibinfo {pages} {1553} (\bibinfo {year}
  {1996})}\BibitemShut {NoStop}%
\bibitem [{\citenamefont {Vitória}\ and\ \citenamefont
  {Belich}(2019)}]{vitoria2019massive}%
  \BibitemOpen
  \bibfield  {author} {\bibinfo {author} {\bibfnamefont {R.~L.~L.}\
  \bibnamefont {Vitória}}\ and\ \bibinfo {author} {\bibfnamefont
  {H.}~\bibnamefont {Belich}},\ }\bibfield  {journal} {\bibinfo  {journal}
  {Advances in High Energy Physics}\ }\href
  {https://doi.org/10.1155/2019/8462973} {10.1155/2019/8462973} (\bibinfo
  {year} {2019})\BibitemShut {NoStop}%
\bibitem [{\citenamefont {Elliott}\ \emph {et~al.}(1994)\citenamefont
  {Elliott}, \citenamefont {King},\ and\ \citenamefont
  {White}}]{king1994radiative}%
  \BibitemOpen
  \bibfield  {author} {\bibinfo {author} {\bibfnamefont {T.}~\bibnamefont
  {Elliott}}, \bibinfo {author} {\bibfnamefont {S.~F.}\ \bibnamefont {King}},\
  and\ \bibinfo {author} {\bibfnamefont {P.~L.}\ \bibnamefont {White}},\ }\href
  {https://doi.org/10.1103/PhysRevD.49.2435} {\bibfield  {journal} {\bibinfo
  {journal} {Physical Review D}\ }\textbf {\bibinfo {volume} {49}},\ \bibinfo
  {pages} {2435} (\bibinfo {year} {1994})}\BibitemShut {NoStop}%
\bibitem [{\citenamefont {Mattingly}(2005)}]{Mattingly2005}%
  \BibitemOpen
  \bibfield  {author} {\bibinfo {author} {\bibfnamefont {D.}~\bibnamefont
  {Mattingly}},\ }\href {https://doi.org/https://doi.org/10.12942/lrr-2005-5}
  {\bibfield  {journal} {\bibinfo  {journal} {Living Rev. Relativity}\ }\textbf
  {\bibinfo {volume} {8}},\ \bibinfo {pages} {5} (\bibinfo {year}
  {2005})}\BibitemShut {NoStop}%
\bibitem [{\citenamefont {Amelino-Camelia}(2013)}]{Amelino-Camelia2013}%
  \BibitemOpen
  \bibfield  {author} {\bibinfo {author} {\bibfnamefont {G.}~\bibnamefont
  {Amelino-Camelia}},\ }\href
  {https://doi.org/https://doi.org/10.12942/lrr-2013-5} {\bibfield  {journal}
  {\bibinfo  {journal} {Living Rev. Relativity}\ }\textbf {\bibinfo {volume}
  {16}},\ \bibinfo {pages} {5} (\bibinfo {year} {2013})}\BibitemShut {NoStop}%
\bibitem [{\citenamefont {Fumeron}(2022)}]{fumeron2022transport}%
  \BibitemOpen
  \bibfield  {author} {\bibinfo {author} {\bibfnamefont {S.}~\bibnamefont
  {Fumeron}},\ }\href {https://hal.science/tel-03607594/document} {\bibfield
  {journal} {\bibinfo  {journal} {HAL Open Science}\ } (\bibinfo {year}
  {2022})}\BibitemShut {NoStop}%
\bibitem [{\citenamefont {Akhmerov}(2011)}]{akhmerov2011dirac}%
  \BibitemOpen
  \bibfield  {author} {\bibinfo {author} {\bibfnamefont {A.~R.}\ \bibnamefont
  {Akhmerov}},\ }\href
  {https://ilorentz.org/beenakkr/mesoscopics/theses/akhmerov/akhmerov.pdf}
  {\bibfield  {journal} {\bibinfo  {journal} {Casimir PhD Series}\ } (\bibinfo
  {year} {2011})}\BibitemShut {NoStop}%
\bibitem [{\citenamefont {Lepori}\ and\ \citenamefont
  {Mussardo}(2010)}]{lepori2010qft}%
  \BibitemOpen
  \bibfield  {author} {\bibinfo {author} {\bibfnamefont {G.}~\bibnamefont
  {Lepori}}\ and\ \bibinfo {author} {\bibfnamefont {G.}~\bibnamefont
  {Mussardo}},\ }\href {https://www.statphys.sissa.it/tesi_phd/lepori.pdf}
  {\bibfield  {journal} {\bibinfo  {journal} {SISSA Statphys}\ } (\bibinfo
  {year} {2010})}\BibitemShut {NoStop}%
\bibitem [{\citenamefont {Ferreira}\ \emph {et~al.}(2017)\citenamefont
  {Ferreira}, \citenamefont {Godinho},\ and\ \citenamefont
  {Neto}}]{ferreira2016torsion}%
  \BibitemOpen
  \bibfield  {author} {\bibinfo {author} {\bibfnamefont {C.~N.}\ \bibnamefont
  {Ferreira}}, \bibinfo {author} {\bibfnamefont {C.~F.~L.}\ \bibnamefont
  {Godinho}},\ and\ \bibinfo {author} {\bibfnamefont {J.~A.~H.}\ \bibnamefont
  {Neto}},\ }\href {https://doi.org/10.1002/andp.201600186} {\bibfield
  {journal} {\bibinfo  {journal} {Annalen der Physik}\ }\textbf {\bibinfo
  {volume} {529}},\ \bibinfo {pages} {1600186} (\bibinfo {year}
  {2017})}\BibitemShut {NoStop}%
\bibitem [{\citenamefont {Abreu}\ \emph {et~al.}(2013)\citenamefont {Abreu},
  \citenamefont {Andrade},\ and\ \citenamefont {Assis}}]{abreu2013vortex}%
  \BibitemOpen
  \bibfield  {author} {\bibinfo {author} {\bibfnamefont {E.}~\bibnamefont
  {Abreu}}, \bibinfo {author} {\bibfnamefont {M.~A.~d.}\ \bibnamefont
  {Andrade}},\ and\ \bibinfo {author} {\bibfnamefont {L.~P. G.~d.}\
  \bibnamefont {Assis}},\ }\href {https://arxiv.org/pdf/1308.2028} {\bibfield
  {journal} {\bibinfo  {journal} {arXiv preprint arXiv:1308.2028}\ } (\bibinfo
  {year} {2013})}\BibitemShut {NoStop}%
\bibitem [{\citenamefont {McKay}\ \emph {et~al.}(2023)\citenamefont {McKay},
  \citenamefont {Mahmood},\ and\ \citenamefont
  {Bradlyn}}]{mckay2023electromagnetic}%
  \BibitemOpen
  \bibfield  {author} {\bibinfo {author} {\bibfnamefont {R.~C.}\ \bibnamefont
  {McKay}}, \bibinfo {author} {\bibfnamefont {F.}~\bibnamefont {Mahmood}},\
  and\ \bibinfo {author} {\bibfnamefont {B.}~\bibnamefont {Bradlyn}},\ }\href
  {https://arxiv.org/abs/2309.11658} {\bibfield  {journal} {\bibinfo  {journal}
  {arXiv preprint arXiv:2309.11658}\ } (\bibinfo {year} {2023})}\BibitemShut
  {NoStop}%
\bibitem [{\citenamefont {Essig}\ \emph {et~al.}(2022)\citenamefont {Essig},
  \citenamefont {Kahn}, \citenamefont {Knapen},\ and\ \citenamefont
  {Ringwald}}]{essig2022snowmass}%
  \BibitemOpen
  \bibfield  {author} {\bibinfo {author} {\bibfnamefont {R.}~\bibnamefont
  {Essig}}, \bibinfo {author} {\bibfnamefont {Y.}~\bibnamefont {Kahn}},
  \bibinfo {author} {\bibfnamefont {S.}~\bibnamefont {Knapen}},\ and\ \bibinfo
  {author} {\bibfnamefont {A.}~\bibnamefont {Ringwald}},\ }\href
  {https://arxiv.org/abs/2203.10089} {\bibfield  {journal} {\bibinfo  {journal}
  {arXiv preprint arXiv:2203.10089}\ } (\bibinfo {year} {2022})}\BibitemShut
  {NoStop}%
\bibitem [{\citenamefont {Francesco}(2012)}]{francesco2012pregeometry}%
  \BibitemOpen
  \bibfield  {author} {\bibinfo {author} {\bibfnamefont {C.}~\bibnamefont
  {Francesco}},\ }\href {https://uwspace.uwaterloo.ca/handle/10012/6868}
  {\bibfield  {journal} {\bibinfo  {journal} {University of Waterloo Theses}\ }
  (\bibinfo {year} {2012})}\BibitemShut {NoStop}%
\bibitem [{\citenamefont {Cortijo}\ and\ \citenamefont
  {Vozmediano}(2007)}]{cortijo2007electronic}%
  \BibitemOpen
  \bibfield  {author} {\bibinfo {author} {\bibfnamefont {A.}~\bibnamefont
  {Cortijo}}\ and\ \bibinfo {author} {\bibfnamefont {M.~A.~H.}\ \bibnamefont
  {Vozmediano}},\ }\href {https://doi.org/10.1016/j.nuclphysb.2006.11.017}
  {\bibfield  {journal} {\bibinfo  {journal} {Nuclear Physics B}\ }\textbf
  {\bibinfo {volume} {763}},\ \bibinfo {pages} {293} (\bibinfo {year}
  {2007})}\BibitemShut {NoStop}%
\bibitem [{\citenamefont {Chamon}\ \emph {et~al.}(2008)\citenamefont {Chamon},
  \citenamefont {Jackiw}, \citenamefont {Pi},\ and\ \citenamefont
  {Santos}}]{chamon2008geometric}%
  \BibitemOpen
  \bibfield  {author} {\bibinfo {author} {\bibfnamefont {C.}~\bibnamefont
  {Chamon}}, \bibinfo {author} {\bibfnamefont {R.}~\bibnamefont {Jackiw}},
  \bibinfo {author} {\bibfnamefont {S.-Y.}\ \bibnamefont {Pi}},\ and\ \bibinfo
  {author} {\bibfnamefont {L.}~\bibnamefont {Santos}},\ }\href
  {https://doi.org/10.1103/PhysRevB.77.235431} {\bibfield  {journal} {\bibinfo
  {journal} {Physical Review B}\ }\textbf {\bibinfo {volume} {77}},\ \bibinfo
  {pages} {235431} (\bibinfo {year} {2008})}\BibitemShut {NoStop}%
\bibitem [{\citenamefont {Garcia~de
  Andrade}(2022)}]{garciadeandrade2022generation}%
  \BibitemOpen
  \bibfield  {author} {\bibinfo {author} {\bibfnamefont {L.~C.}\ \bibnamefont
  {Garcia~de Andrade}},\ }\href {https://doi.org/10.3390/universe8120658}
  {\bibfield  {journal} {\bibinfo  {journal} {Universe}\ }\textbf {\bibinfo
  {volume} {8}},\ \bibinfo {pages} {658} (\bibinfo {year} {2022})}\BibitemShut
  {NoStop}%
\bibitem [{\citenamefont {Käding}\ \emph {et~al.}(2019)\citenamefont
  {Käding}, \citenamefont {Millington}, \citenamefont {Minář},\ and\
  \citenamefont {Burrage}}]{kaeding2019open}%
  \BibitemOpen
  \bibfield  {author} {\bibinfo {author} {\bibfnamefont {C.}~\bibnamefont
  {Käding}}, \bibinfo {author} {\bibfnamefont {P.}~\bibnamefont {Millington}},
  \bibinfo {author} {\bibfnamefont {J.}~\bibnamefont {Minář}},\ and\ \bibinfo
  {author} {\bibfnamefont {C.}~\bibnamefont {Burrage}},\ }\href
  {https://doi.org/10.1103/PhysRevD.100.076003} {\bibfield  {journal} {\bibinfo
   {journal} {Physical Review D}\ }\textbf {\bibinfo {volume} {100}},\ \bibinfo
  {pages} {076003} (\bibinfo {year} {2019})}\BibitemShut {NoStop}%
\bibitem [{\citenamefont {Bakke}\ \emph {et~al.}(2012)\citenamefont {Bakke},
  \citenamefont {Silva},\ and\ \citenamefont {Belich}}]{bakke2012effect}%
  \BibitemOpen
  \bibfield  {author} {\bibinfo {author} {\bibfnamefont {K.}~\bibnamefont
  {Bakke}}, \bibinfo {author} {\bibfnamefont {E.~O.}\ \bibnamefont {Silva}},\
  and\ \bibinfo {author} {\bibfnamefont {H.}~\bibnamefont {Belich}},\ }\href
  {https://doi.org/10.1088/0954-3899/39/5/055004} {\bibfield  {journal}
  {\bibinfo  {journal} {Journal of Physics G: Nuclear and Particle Physics}\
  }\textbf {\bibinfo {volume} {39}},\ \bibinfo {pages} {055004} (\bibinfo
  {year} {2012})}\BibitemShut {NoStop}%
\bibitem [{\citenamefont {Dalzell}\ \emph {et~al.}(2023)\citenamefont
  {Dalzell}, \citenamefont {McArdle}, \citenamefont {Berta},\ and\
  \citenamefont {Bienias}}]{dalzell2023survey}%
  \BibitemOpen
  \bibfield  {author} {\bibinfo {author} {\bibfnamefont {A.~M.}\ \bibnamefont
  {Dalzell}}, \bibinfo {author} {\bibfnamefont {S.}~\bibnamefont {McArdle}},
  \bibinfo {author} {\bibfnamefont {M.}~\bibnamefont {Berta}},\ and\ \bibinfo
  {author} {\bibfnamefont {P.}~\bibnamefont {Bienias}},\ }\href
  {https://arxiv.org/pdf/2310.03011} {\bibfield  {journal} {\bibinfo  {journal}
  {arXiv preprint arXiv:2310.03011}\ } (\bibinfo {year} {2023})}\BibitemShut
  {NoStop}%
\bibitem [{\citenamefont {Alexandre}\ and\ \citenamefont
  {Leite}(2016)}]{alexandre2016kinematics}%
  \BibitemOpen
  \bibfield  {author} {\bibinfo {author} {\bibfnamefont {J.}~\bibnamefont
  {Alexandre}}\ and\ \bibinfo {author} {\bibfnamefont {J.}~\bibnamefont
  {Leite}},\ }\href {https://doi.org/10.1088/0264-9381/33/19/195005} {\bibfield
   {journal} {\bibinfo  {journal} {Classical and Quantum Gravity}\ }\textbf
  {\bibinfo {volume} {33}},\ \bibinfo {pages} {195005} (\bibinfo {year}
  {2016})}\BibitemShut {NoStop}%
\bibitem [{\citenamefont {Morales-Pérez}\ \emph {et~al.}(2023)\citenamefont
  {Morales-Pérez}, \citenamefont {Devescovi},\ and\ \citenamefont
  {Hwang}}]{moralesperez2023tightbinding}%
  \BibitemOpen
  \bibfield  {author} {\bibinfo {author} {\bibfnamefont {A.}~\bibnamefont
  {Morales-Pérez}}, \bibinfo {author} {\bibfnamefont {C.}~\bibnamefont
  {Devescovi}},\ and\ \bibinfo {author} {\bibfnamefont {Y.}~\bibnamefont
  {Hwang}},\ }\href {https://arxiv.org/abs/2305.18257} {\bibfield  {journal}
  {\bibinfo  {journal} {arXiv preprint arXiv:2305.18257}\ } (\bibinfo {year}
  {2023})}\BibitemShut {NoStop}%
\bibitem [{\citenamefont {Candelas}\ and\ \citenamefont
  {Weinberg}(1984)}]{candelas1984gauge}%
  \BibitemOpen
  \bibfield  {author} {\bibinfo {author} {\bibfnamefont {P.}~\bibnamefont
  {Candelas}}\ and\ \bibinfo {author} {\bibfnamefont {S.}~\bibnamefont
  {Weinberg}},\ }\href {https://doi.org/10.1016/0550-3213(84)90001-4}
  {\bibfield  {journal} {\bibinfo  {journal} {Nuclear Physics B}\ }\textbf
  {\bibinfo {volume} {255}},\ \bibinfo {pages} {189} (\bibinfo {year}
  {1984})}\BibitemShut {NoStop}%
\bibitem [{\citenamefont {Ferreira~Jr.}\ \emph {et~al.}(2019)\citenamefont
  {Ferreira~Jr.}, \citenamefont {Reis},\ and\ \citenamefont
  {Schreck}}]{ferreira2019dimensional}%
  \BibitemOpen
  \bibfield  {author} {\bibinfo {author} {\bibfnamefont {M.~M.}\ \bibnamefont
  {Ferreira~Jr.}}, \bibinfo {author} {\bibfnamefont {J.~A. A.~S.}\ \bibnamefont
  {Reis}},\ and\ \bibinfo {author} {\bibfnamefont {M.}~\bibnamefont
  {Schreck}},\ }\href {https://doi.org/10.1103/PhysRevD.100.095026} {\bibfield
  {journal} {\bibinfo  {journal} {Physical Review D}\ }\textbf {\bibinfo
  {volume} {100}},\ \bibinfo {pages} {095026} (\bibinfo {year}
  {2019})}\BibitemShut {NoStop}%
\bibitem [{\citenamefont {Lisboa-Santos}\ \emph
  {et~al.}(2023{\natexlab{b}})\citenamefont {Lisboa-Santos}, \citenamefont
  {Reis},\ and\ \citenamefont {Schreck}}]{lisboa2023planar}%
  \BibitemOpen
  \bibfield  {author} {\bibinfo {author} {\bibfnamefont {L.}~\bibnamefont
  {Lisboa-Santos}}, \bibinfo {author} {\bibfnamefont {J.~A. A.~S.}\
  \bibnamefont {Reis}},\ and\ \bibinfo {author} {\bibfnamefont
  {M.}~\bibnamefont {Schreck}},\ }\href
  {https://doi.org/10.1103/PhysRevD.108.115032} {\bibfield  {journal} {\bibinfo
   {journal} {Physical Review D}\ }\textbf {\bibinfo {volume} {108}},\ \bibinfo
  {pages} {115032} (\bibinfo {year} {2023}{\natexlab{b}})}\BibitemShut
  {NoStop}%
\bibitem [{\citenamefont {Ferrari}\ \emph {et~al.}(2024)\citenamefont
  {Ferrari}, \citenamefont {Rocha},\ and\ \citenamefont
  {Borges}}]{ferrari2024interactions}%
  \BibitemOpen
  \bibfield  {author} {\bibinfo {author} {\bibfnamefont {A.~F.}\ \bibnamefont
  {Ferrari}}, \bibinfo {author} {\bibfnamefont {C.~B.}\ \bibnamefont {Rocha}},\
  and\ \bibinfo {author} {\bibfnamefont {L.~H.~C.}\ \bibnamefont {Borges}},\
  }\href {https://link.springer.com/article/10.1007/s13538-024-01598-5}
  {\bibfield  {journal} {\bibinfo  {journal} {Brazilian Journal of Physics}\ }
  (\bibinfo {year} {2024})}\BibitemShut {NoStop}%
\bibitem [{\citenamefont {Liu}\ \emph {et~al.}(2016)\citenamefont {Liu},
  \citenamefont {Zhang},\ and\ \citenamefont {Qi}}]{liu2016quantum}%
  \BibitemOpen
  \bibfield  {author} {\bibinfo {author} {\bibfnamefont {C.-X.}\ \bibnamefont
  {Liu}}, \bibinfo {author} {\bibfnamefont {S.-C.}\ \bibnamefont {Zhang}},\
  and\ \bibinfo {author} {\bibfnamefont {X.-L.}\ \bibnamefont {Qi}},\ }\href
  {https://doi.org/10.1146/annurev-conmatphys-031115-011417} {\bibfield
  {journal} {\bibinfo  {journal} {Annual Review of Condensed Matter Physics}\
  }\textbf {\bibinfo {volume} {7}},\ \bibinfo {pages} {301} (\bibinfo {year}
  {2016})}\BibitemShut {NoStop}%
\bibitem [{\citenamefont {Adamo}\ \emph {et~al.}(2009)\citenamefont {Adamo},
  \citenamefont {MacDonald}, \citenamefont {Fu}, \citenamefont {Wang},
  \citenamefont {Tsai},\ and\ \citenamefont {Zheludev}}]{adamo2009light}%
  \BibitemOpen
  \bibfield  {author} {\bibinfo {author} {\bibfnamefont {G.}~\bibnamefont
  {Adamo}}, \bibinfo {author} {\bibfnamefont {K.~F.}\ \bibnamefont
  {MacDonald}}, \bibinfo {author} {\bibfnamefont {Y.~H.}\ \bibnamefont {Fu}},
  \bibinfo {author} {\bibfnamefont {C.~M.}\ \bibnamefont {Wang}}, \bibinfo
  {author} {\bibfnamefont {D.~P.}\ \bibnamefont {Tsai}},\ and\ \bibinfo
  {author} {\bibfnamefont {N.~I.}\ \bibnamefont {Zheludev}},\ }\href
  {https://doi.org/10.1103/PhysRevLett.103.113901} {\bibfield  {journal}
  {\bibinfo  {journal} {Physical Review Letters}\ }\textbf {\bibinfo {volume}
  {103}},\ \bibinfo {pages} {113901} (\bibinfo {year} {2009})}\BibitemShut
  {NoStop}%
\bibitem [{\citenamefont {Chang}\ \emph {et~al.}(2006)\citenamefont {Chang},
  \citenamefont {Sørensen}, \citenamefont {Hemmer},\ and\ \citenamefont
  {Lukin}}]{chang2006quantum}%
  \BibitemOpen
  \bibfield  {author} {\bibinfo {author} {\bibfnamefont {D.~E.}\ \bibnamefont
  {Chang}}, \bibinfo {author} {\bibfnamefont {A.~S.}\ \bibnamefont
  {Sørensen}}, \bibinfo {author} {\bibfnamefont {P.~R.}\ \bibnamefont
  {Hemmer}},\ and\ \bibinfo {author} {\bibfnamefont {M.~D.}\ \bibnamefont
  {Lukin}},\ }\href {https://doi.org/10.1103/PhysRevLett.97.053002} {\bibfield
  {journal} {\bibinfo  {journal} {Physical Review Letters}\ }\textbf {\bibinfo
  {volume} {97}},\ \bibinfo {pages} {053002} (\bibinfo {year}
  {2006})}\BibitemShut {NoStop}%
\bibitem [{\citenamefont {McKay}\ \emph {et~al.}(2024)\citenamefont {McKay},
  \citenamefont {Mahmood},\ and\ \citenamefont {Bradlyn}}]{mckay2024spatially}%
  \BibitemOpen
  \bibfield  {author} {\bibinfo {author} {\bibfnamefont {R.~C.}\ \bibnamefont
  {McKay}}, \bibinfo {author} {\bibfnamefont {F.}~\bibnamefont {Mahmood}},\
  and\ \bibinfo {author} {\bibfnamefont {B.}~\bibnamefont {Bradlyn}},\ }\href
  {https://doi.org/10.1103/PhysRevX.14.011058} {\bibfield  {journal} {\bibinfo
  {journal} {Physical Review X}\ }\textbf {\bibinfo {volume} {14}},\ \bibinfo
  {pages} {011058} (\bibinfo {year} {2024})}\BibitemShut {NoStop}%
\bibitem [{\citenamefont {Barontini}\ \emph {et~al.}(2022)\citenamefont
  {Barontini}, \citenamefont {Blackburn},\ and\ \citenamefont
  {Boyer}}]{barontini2022measuring}%
  \BibitemOpen
  \bibfield  {author} {\bibinfo {author} {\bibfnamefont {G.}~\bibnamefont
  {Barontini}}, \bibinfo {author} {\bibfnamefont {L.}~\bibnamefont
  {Blackburn}},\ and\ \bibinfo {author} {\bibfnamefont {V.~e.~a.}\ \bibnamefont
  {Boyer}},\ }\href {https://doi.org/10.1140/epjqt/s40507-022-00130-5}
  {\bibfield  {journal} {\bibinfo  {journal} {EPJ Quantum Technology}\ }\textbf
  {\bibinfo {volume} {9}},\ \bibinfo {pages} {47} (\bibinfo {year}
  {2022})}\BibitemShut {NoStop}%
\bibitem [{\citenamefont {Arias}\ \emph {et~al.}(2012)\citenamefont {Arias},
  \citenamefont {Cadamuro},\ and\ \citenamefont {Goodsell}}]{arias2012wispy}%
  \BibitemOpen
  \bibfield  {author} {\bibinfo {author} {\bibfnamefont {P.}~\bibnamefont
  {Arias}}, \bibinfo {author} {\bibfnamefont {D.}~\bibnamefont {Cadamuro}},\
  and\ \bibinfo {author} {\bibfnamefont {M.~e.~a.}\ \bibnamefont {Goodsell}},\
  }\href {https://doi.org/10.1088/1475-7516/2012/06/013} {\bibfield  {journal}
  {\bibinfo  {journal} {Journal of Cosmology and Astroparticle Physics}\
  }\textbf {\bibinfo {volume} {2012}},\ \bibinfo {pages} {013}}\BibitemShut
  {NoStop}%
\bibitem [{\citenamefont {Crescini}(2019)}]{cresini2019towards}%
  \BibitemOpen
  \bibfield  {author} {\bibinfo {author} {\bibfnamefont {N.}~\bibnamefont
  {Crescini}},\ }\href {https://www.research.unipd.it/handle/11577/3425918}
  {\bibfield  {journal} {\bibinfo  {journal} {Research Unipd}\ } (\bibinfo
  {year} {2019})}\BibitemShut {NoStop}%
\bibitem [{\citenamefont {Reidler}\ and\ \citenamefont
  {Yelin}(2008)}]{reidler2008quantum}%
  \BibitemOpen
  \bibfield  {author} {\bibinfo {author} {\bibfnamefont {H.}~\bibnamefont
  {Reidler}}\ and\ \bibinfo {author} {\bibfnamefont {D.}~\bibnamefont
  {Yelin}},\ }\href {https://doi.org/10.1109/JQE.2007.912734} {\bibfield
  {journal} {\bibinfo  {journal} {Journal of Quantum Electronics}\ }\textbf
  {\bibinfo {volume} {44}},\ \bibinfo {pages} {145} (\bibinfo {year}
  {2008})}\BibitemShut {NoStop}%
\bibitem [{\citenamefont {Xia}\ \emph {et~al.}(2009)\citenamefont {Xia},
  \citenamefont {Qian}, \citenamefont {Hsieh}, \citenamefont {Wray},
  \citenamefont {Pal}, \citenamefont {Lin}, \citenamefont {Bansil},
  \citenamefont {Grauer}, \citenamefont {Hor}, \citenamefont {Cava},\ and\
  \citenamefont {Hasan}}]{xia2009large}%
  \BibitemOpen
  \bibfield  {author} {\bibinfo {author} {\bibfnamefont {Y.}~\bibnamefont
  {Xia}}, \bibinfo {author} {\bibfnamefont {D.}~\bibnamefont {Qian}}, \bibinfo
  {author} {\bibfnamefont {D.}~\bibnamefont {Hsieh}}, \bibinfo {author}
  {\bibfnamefont {L.}~\bibnamefont {Wray}}, \bibinfo {author} {\bibfnamefont
  {A.}~\bibnamefont {Pal}}, \bibinfo {author} {\bibfnamefont {H.}~\bibnamefont
  {Lin}}, \bibinfo {author} {\bibfnamefont {A.}~\bibnamefont {Bansil}},
  \bibinfo {author} {\bibfnamefont {D.}~\bibnamefont {Grauer}}, \bibinfo
  {author} {\bibfnamefont {Y.~S.}\ \bibnamefont {Hor}}, \bibinfo {author}
  {\bibfnamefont {R.~J.}\ \bibnamefont {Cava}},\ and\ \bibinfo {author}
  {\bibfnamefont {M.~Z.}\ \bibnamefont {Hasan}},\ }\href
  {https://doi.org/10.1038/nphys1274} {\bibfield  {journal} {\bibinfo
  {journal} {Nature Physics}\ }\textbf {\bibinfo {volume} {5}},\ \bibinfo
  {pages} {398} (\bibinfo {year} {2009})}\BibitemShut {NoStop}%
\bibitem [{\citenamefont {Burkov}(2015)}]{burkov2015chiral}%
  \BibitemOpen
  \bibfield  {author} {\bibinfo {author} {\bibfnamefont {A.~A.}\ \bibnamefont
  {Burkov}},\ }\href {https://doi.org/10.1088/0953-8984/27/11/113201}
  {\bibfield  {journal} {\bibinfo  {journal} {Journal of Physics: Condensed
  Matter}\ }\textbf {\bibinfo {volume} {27}},\ \bibinfo {pages} {113201}
  (\bibinfo {year} {2015})}\BibitemShut {NoStop}%
\bibitem [{\citenamefont {Li}\ \emph {et~al.}(2016)\citenamefont {Li},
  \citenamefont {Kharzeev}, \citenamefont {Zhang}, \citenamefont {Huang},
  \citenamefont {Pletikosic}, \citenamefont {Fedorov}, \citenamefont {Zhong},
  \citenamefont {Schneeloch}, \citenamefont {Gu},\ and\ \citenamefont
  {Valla}}]{li2016chiral}%
  \BibitemOpen
  \bibfield  {author} {\bibinfo {author} {\bibfnamefont {Q.}~\bibnamefont
  {Li}}, \bibinfo {author} {\bibfnamefont {D.~E.}\ \bibnamefont {Kharzeev}},
  \bibinfo {author} {\bibfnamefont {C.}~\bibnamefont {Zhang}}, \bibinfo
  {author} {\bibfnamefont {Y.}~\bibnamefont {Huang}}, \bibinfo {author}
  {\bibfnamefont {I.}~\bibnamefont {Pletikosic}}, \bibinfo {author}
  {\bibfnamefont {A.~V.}\ \bibnamefont {Fedorov}}, \bibinfo {author}
  {\bibfnamefont {R.~D.}\ \bibnamefont {Zhong}}, \bibinfo {author}
  {\bibfnamefont {J.}~\bibnamefont {Schneeloch}}, \bibinfo {author}
  {\bibfnamefont {G.~D.}\ \bibnamefont {Gu}},\ and\ \bibinfo {author}
  {\bibfnamefont {T.}~\bibnamefont {Valla}},\ }\href
  {https://doi.org/10.1038/nphys3648} {\bibfield  {journal} {\bibinfo
  {journal} {Nature Physics}\ }\textbf {\bibinfo {volume} {12}},\ \bibinfo
  {pages} {550} (\bibinfo {year} {2016})}\BibitemShut {NoStop}%
\bibitem [{\citenamefont {Potter}\ \emph {et~al.}(2014)\citenamefont {Potter},
  \citenamefont {Kimchi},\ and\ \citenamefont
  {Vishwanath}}]{potter2014quantum}%
  \BibitemOpen
  \bibfield  {author} {\bibinfo {author} {\bibfnamefont {A.~C.}\ \bibnamefont
  {Potter}}, \bibinfo {author} {\bibfnamefont {I.}~\bibnamefont {Kimchi}},\
  and\ \bibinfo {author} {\bibfnamefont {A.}~\bibnamefont {Vishwanath}},\
  }\href {https://doi.org/10.1038/ncomms6161} {\bibfield  {journal} {\bibinfo
  {journal} {Nature Communications}\ }\textbf {\bibinfo {volume} {5}},\
  \bibinfo {pages} {5161} (\bibinfo {year} {2014})}\BibitemShut {NoStop}%
\bibitem [{\citenamefont {Huang}\ \emph {et~al.}(2015)\citenamefont {Huang},
  \citenamefont {Zhao}, \citenamefont {Long}, \citenamefont {Wang},
  \citenamefont {Chen}, \citenamefont {Yang}, \citenamefont {Liang},
  \citenamefont {Xue}, \citenamefont {Weng}, \citenamefont {Fang},
  \citenamefont {Dai},\ and\ \citenamefont {Chen}}]{huang2015observation}%
  \BibitemOpen
  \bibfield  {author} {\bibinfo {author} {\bibfnamefont {X.}~\bibnamefont
  {Huang}}, \bibinfo {author} {\bibfnamefont {L.}~\bibnamefont {Zhao}},
  \bibinfo {author} {\bibfnamefont {Y.}~\bibnamefont {Long}}, \bibinfo {author}
  {\bibfnamefont {P.}~\bibnamefont {Wang}}, \bibinfo {author} {\bibfnamefont
  {D.}~\bibnamefont {Chen}}, \bibinfo {author} {\bibfnamefont {Z.}~\bibnamefont
  {Yang}}, \bibinfo {author} {\bibfnamefont {H.}~\bibnamefont {Liang}},
  \bibinfo {author} {\bibfnamefont {M.}~\bibnamefont {Xue}}, \bibinfo {author}
  {\bibfnamefont {H.}~\bibnamefont {Weng}}, \bibinfo {author} {\bibfnamefont
  {Z.}~\bibnamefont {Fang}}, \bibinfo {author} {\bibfnamefont {X.}~\bibnamefont
  {Dai}},\ and\ \bibinfo {author} {\bibfnamefont {G.}~\bibnamefont {Chen}},\
  }\href {https://doi.org/10.1103/PhysRevX.5.031023} {\bibfield  {journal}
  {\bibinfo  {journal} {Physical Review X}\ }\textbf {\bibinfo {volume} {5}},\
  \bibinfo {pages} {031023} (\bibinfo {year} {2015})}\BibitemShut {NoStop}%
\bibitem [{\citenamefont {Deng}\ \emph {et~al.}(2021)\citenamefont {Deng},
  \citenamefont {Van~Dyke}, \citenamefont {Minic}, \citenamefont {Heremans},\
  and\ \citenamefont {Barnes}}]{PhysRevB.104.075202}%
  \BibitemOpen
  \bibfield  {author} {\bibinfo {author} {\bibfnamefont {K.}~\bibnamefont
  {Deng}}, \bibinfo {author} {\bibfnamefont {J.~S.}\ \bibnamefont {Van~Dyke}},
  \bibinfo {author} {\bibfnamefont {D.}~\bibnamefont {Minic}}, \bibinfo
  {author} {\bibfnamefont {J.~J.}\ \bibnamefont {Heremans}},\ and\ \bibinfo
  {author} {\bibfnamefont {E.}~\bibnamefont {Barnes}},\ }\href
  {https://doi.org/10.1103/PhysRevB.104.075202} {\bibfield  {journal} {\bibinfo
   {journal} {Phys. Rev. B}\ }\textbf {\bibinfo {volume} {104}},\ \bibinfo
  {pages} {075202} (\bibinfo {year} {2021})}\BibitemShut {NoStop}%
\bibitem [{\citenamefont {Kane}\ and\ \citenamefont
  {Mele}(2005)}]{kane2005graphene}%
  \BibitemOpen
  \bibfield  {author} {\bibinfo {author} {\bibfnamefont {C.~L.}\ \bibnamefont
  {Kane}}\ and\ \bibinfo {author} {\bibfnamefont {E.~J.}\ \bibnamefont
  {Mele}},\ }\href {https://doi.org/10.1103/PhysRevLett.95.226801} {\bibfield
  {journal} {\bibinfo  {journal} {Physical Review Letters}\ }\textbf {\bibinfo
  {volume} {95}},\ \bibinfo {pages} {226801} (\bibinfo {year}
  {2005})}\BibitemShut {NoStop}%
\bibitem [{\citenamefont {Sitte}\ \emph {et~al.}(2012)\citenamefont {Sitte},
  \citenamefont {Rosch}, \citenamefont {Altman},\ and\ \citenamefont
  {Fritz}}]{sitte2012topological}%
  \BibitemOpen
  \bibfield  {author} {\bibinfo {author} {\bibfnamefont {M.}~\bibnamefont
  {Sitte}}, \bibinfo {author} {\bibfnamefont {A.}~\bibnamefont {Rosch}},
  \bibinfo {author} {\bibfnamefont {E.}~\bibnamefont {Altman}},\ and\ \bibinfo
  {author} {\bibfnamefont {L.}~\bibnamefont {Fritz}},\ }\href
  {https://doi.org/10.1103/PhysRevLett.108.126807} {\bibfield  {journal}
  {\bibinfo  {journal} {Physical Review Letters}\ }\textbf {\bibinfo {volume}
  {108}},\ \bibinfo {pages} {126807} (\bibinfo {year} {2012})}\BibitemShut
  {NoStop}%
\bibitem [{\citenamefont {Kolovsky}(2011)}]{kolovsky2011hall}%
  \BibitemOpen
  \bibfield  {author} {\bibinfo {author} {\bibfnamefont {A.~R.}\ \bibnamefont
  {Kolovsky}},\ }\href {https://doi.org/10.1209/0295-5075/96/50002} {\bibfield
  {journal} {\bibinfo  {journal} {Europhysics Letters}\ }\textbf {\bibinfo
  {volume} {96}},\ \bibinfo {pages} {50002} (\bibinfo {year}
  {2011})}\BibitemShut {NoStop}%
\bibitem [{\citenamefont {Sharma}\ \emph {et~al.}(2016)\citenamefont {Sharma},
  \citenamefont {Goswami},\ and\ \citenamefont {Tewari}}]{sharma2016nernst}%
  \BibitemOpen
  \bibfield  {author} {\bibinfo {author} {\bibfnamefont {G.}~\bibnamefont
  {Sharma}}, \bibinfo {author} {\bibfnamefont {P.}~\bibnamefont {Goswami}},\
  and\ \bibinfo {author} {\bibfnamefont {S.}~\bibnamefont {Tewari}},\ }\href
  {https://doi.org/10.1103/PhysRevB.93.035116} {\bibfield  {journal} {\bibinfo
  {journal} {Physical Review B}\ }\textbf {\bibinfo {volume} {93}},\ \bibinfo
  {pages} {035116} (\bibinfo {year} {2016})}\BibitemShut {NoStop}%
\bibitem [{\citenamefont {Ferreiros}\ \emph {et~al.}(2017)\citenamefont
  {Ferreiros}, \citenamefont {Zyuzin},\ and\ \citenamefont
  {Bardarson}}]{ferreiros2017anomalous}%
  \BibitemOpen
  \bibfield  {author} {\bibinfo {author} {\bibfnamefont {Y.}~\bibnamefont
  {Ferreiros}}, \bibinfo {author} {\bibfnamefont {A.~A.}\ \bibnamefont
  {Zyuzin}},\ and\ \bibinfo {author} {\bibfnamefont {J.~H.}\ \bibnamefont
  {Bardarson}},\ }\href {https://doi.org/10.1103/PhysRevB.96.115202} {\bibfield
   {journal} {\bibinfo  {journal} {Physical Review B}\ }\textbf {\bibinfo
  {volume} {96}},\ \bibinfo {pages} {115202} (\bibinfo {year}
  {2017})}\BibitemShut {NoStop}%
\bibitem [{\citenamefont {Markov}\ \emph {et~al.}(2019)\citenamefont {Markov},
  \citenamefont {Rohringer},\ and\ \citenamefont
  {Rubtsov}}]{markov2019robustness}%
  \BibitemOpen
  \bibfield  {author} {\bibinfo {author} {\bibfnamefont {A.~A.}\ \bibnamefont
  {Markov}}, \bibinfo {author} {\bibfnamefont {G.}~\bibnamefont {Rohringer}},\
  and\ \bibinfo {author} {\bibfnamefont {A.~N.}\ \bibnamefont {Rubtsov}},\
  }\href {https://doi.org/10.1103/PhysRevB.100.115102} {\bibfield  {journal}
  {\bibinfo  {journal} {Physical Review B}\ }\textbf {\bibinfo {volume}
  {100}},\ \bibinfo {pages} {115102} (\bibinfo {year} {2019})}\BibitemShut
  {NoStop}%
\bibitem [{\citenamefont {Wang}\ \emph {et~al.}(2022)\citenamefont {Wang},
  \citenamefont {Zhu},\ and\ \citenamefont {Su}}]{wang2022quantum}%
  \BibitemOpen
  \bibfield  {author} {\bibinfo {author} {\bibfnamefont {Y.~D.}\ \bibnamefont
  {Wang}}, \bibinfo {author} {\bibfnamefont {Z.~G.}\ \bibnamefont {Zhu}},\ and\
  \bibinfo {author} {\bibfnamefont {G.}~\bibnamefont {Su}},\ }\href
  {https://doi.org/10.1103/PhysRevB.106.035148} {\bibfield  {journal} {\bibinfo
   {journal} {Physical Review B}\ }\textbf {\bibinfo {volume} {106}},\ \bibinfo
  {pages} {035148} (\bibinfo {year} {2022})}\BibitemShut {NoStop}%
\bibitem [{\citenamefont {Ando}(2015)}]{ando2015valley}%
  \BibitemOpen
  \bibfield  {author} {\bibinfo {author} {\bibfnamefont {T.}~\bibnamefont
  {Ando}},\ }\href {https://doi.org/10.7566/JPSJ.84.114705} {\bibfield
  {journal} {\bibinfo  {journal} {Journal of the Physical Society of Japan}\
  }\textbf {\bibinfo {volume} {84}},\ \bibinfo {pages} {114705} (\bibinfo
  {year} {2015})}\BibitemShut {NoStop}%
\bibitem [{\citenamefont {Ching}\ and\ \citenamefont
  {Xu}(1991)}]{ching1987solid}%
  \BibitemOpen
  \bibfield  {author} {\bibinfo {author} {\bibfnamefont {W.~Y.}\ \bibnamefont
  {Ching}}\ and\ \bibinfo {author} {\bibfnamefont {Y.~N.}\ \bibnamefont {Xu}},\
  }\href
  {http://users.wfu.edu/natalie/s11phy752/lecturenote/Casestudies/BN.1991.PhysRevB.44.7787.pdf}
  {\bibfield  {journal} {\bibinfo  {journal} {Physical Review B}\ }\textbf
  {\bibinfo {volume} {44}},\ \bibinfo {pages} {7787} (\bibinfo {year}
  {1991})}\BibitemShut {NoStop}%
\bibitem [{\citenamefont {Louis}(2002)}]{louis2002density}%
  \BibitemOpen
  \bibfield  {author} {\bibinfo {author} {\bibfnamefont {A.~A.}\ \bibnamefont
  {Louis}},\ }\href {https://doi.org/10.1088/0953-8984/14/40/311} {\bibfield
  {journal} {\bibinfo  {journal} {Journal of Physics: Condensed Matter}\
  }\textbf {\bibinfo {volume} {14}},\ \bibinfo {pages} {1233} (\bibinfo {year}
  {2002})}\BibitemShut {NoStop}%
\bibitem [{\citenamefont {Marder}(2010)}]{marder2010condensed}%
  \BibitemOpen
  \bibfield  {author} {\bibinfo {author} {\bibfnamefont {M.~P.}\ \bibnamefont
  {Marder}},\ }\href {https://books.google.com/books?id=ijloadAt4BQC} {\emph
  {\bibinfo {title} {Condensed Matter Physics}}}\ (\bibinfo  {publisher}
  {Wiley},\ \bibinfo {year} {2010})\BibitemShut {NoStop}%
\bibitem [{\citenamefont {Bruus}\ and\ \citenamefont
  {Flensberg}(2004)}]{bruus2004manybody}%
  \BibitemOpen
  \bibfield  {author} {\bibinfo {author} {\bibfnamefont {H.}~\bibnamefont
  {Bruus}}\ and\ \bibinfo {author} {\bibfnamefont {K.}~\bibnamefont
  {Flensberg}},\ }\href {https://kft.umcs.lublin.pl/doman/mechanika/Bruus.pdf}
  {\emph {\bibinfo {title} {Many-body Quantum Theory in Condensed Matter
  Physics: An Introduction}}}\ (\bibinfo  {publisher} {Oxford University
  Press},\ \bibinfo {year} {2004})\BibitemShut {NoStop}%
\bibitem [{\citenamefont {Cohen-Tannoudji}\ \emph {et~al.}(1977)\citenamefont
  {Cohen-Tannoudji}, \citenamefont {Diu},\ and\ \citenamefont
  {Laloë}}]{cohen1977quantum}%
  \BibitemOpen
  \bibfield  {author} {\bibinfo {author} {\bibfnamefont {C.}~\bibnamefont
  {Cohen-Tannoudji}}, \bibinfo {author} {\bibfnamefont {B.}~\bibnamefont
  {Diu}},\ and\ \bibinfo {author} {\bibfnamefont {F.}~\bibnamefont {Laloë}},\
  }\href
  {https://www.wiley.com/en-us/Quantum+Mechanics%2C+2+Volume+Set-p-9780471164333}
  {\emph {\bibinfo {title} {Quantum Mechanics}}},\ Vol.\ \bibinfo {volume} {1
  \& 2}\ (\bibinfo  {publisher} {Wiley},\ \bibinfo {year} {1977})\BibitemShut
  {NoStop}%
\bibitem [{\citenamefont {Schiff}(1968)}]{schiff1968quantum}%
  \BibitemOpen
  \bibfield  {author} {\bibinfo {author} {\bibfnamefont {L.~I.}\ \bibnamefont
  {Schiff}},\ }\href
  {https://books.google.com/books/about/Quantum_Mechanics.html?id=6DsZAQAAIAAJ}
  {\emph {\bibinfo {title} {Quantum Mechanics}}},\ \bibinfo {edition} {3rd}\
  ed.\ (\bibinfo  {publisher} {McGraw-Hill},\ \bibinfo {year}
  {1968})\BibitemShut {NoStop}%
\bibitem [{\citenamefont {Demtröder}(2010)}]{demtroder2010atoms}%
  \BibitemOpen
  \bibfield  {author} {\bibinfo {author} {\bibfnamefont {W.}~\bibnamefont
  {Demtröder}},\ }\href {https://doi.org/10.1007/978-3-642-10298-1} {\emph
  {\bibinfo {title} {Atoms, Molecules and Photons: An Introduction to Atomic-
  Molecular- and Quantum Physics}}},\ \bibinfo {edition} {2nd}\ ed.\ (\bibinfo
  {publisher} {Springer},\ \bibinfo {year} {2010})\BibitemShut {NoStop}%
\bibitem [{\citenamefont {Jackson}(1998)}]{jackson1998classical}%
  \BibitemOpen
  \bibfield  {author} {\bibinfo {author} {\bibfnamefont {J.~D.}\ \bibnamefont
  {Jackson}},\ }\href
  {https://www.wiley.com/en-us/Classical+Electrodynamics%2C+3rd+Edition-p-9780471309321}
  {\emph {\bibinfo {title} {Classical Electrodynamics}}},\ \bibinfo {edition}
  {3rd}\ ed.\ (\bibinfo  {publisher} {Wiley},\ \bibinfo {year}
  {1998})\BibitemShut {NoStop}%
\bibitem [{\citenamefont {Tipler}\ and\ \citenamefont
  {Llewellyn}(2012)}]{tipler2012modern}%
  \BibitemOpen
  \bibfield  {author} {\bibinfo {author} {\bibfnamefont {P.~A.}\ \bibnamefont
  {Tipler}}\ and\ \bibinfo {author} {\bibfnamefont {R.~A.}\ \bibnamefont
  {Llewellyn}},\ }\href
  {https://www.macmillanlearning.com/college/us/product/Modern-Physics/p/142925078X}
  {\emph {\bibinfo {title} {Modern Physics}}},\ \bibinfo {edition} {6th}\ ed.\
  (\bibinfo  {publisher} {W.H. Freeman},\ \bibinfo {year} {2012})\BibitemShut
  {NoStop}%
\bibitem [{\citenamefont {Friedrich}(2005)}]{friedrich2005theoretical}%
  \BibitemOpen
  \bibfield  {author} {\bibinfo {author} {\bibfnamefont {H.}~\bibnamefont
  {Friedrich}},\ }\href {https://doi.org/10.1007/b137686} {\emph {\bibinfo
  {title} {Theoretical Atomic Physics}}},\ \bibinfo {edition} {3rd}\ ed.\
  (\bibinfo  {publisher} {Springer},\ \bibinfo {year} {2005})\BibitemShut
  {NoStop}%
\bibitem [{\citenamefont {Scully}\ and\ \citenamefont
  {Zubairy}(1997)}]{scully1997quantum}%
  \BibitemOpen
  \bibfield  {author} {\bibinfo {author} {\bibfnamefont {M.~O.}\ \bibnamefont
  {Scully}}\ and\ \bibinfo {author} {\bibfnamefont {M.~S.}\ \bibnamefont
  {Zubairy}},\ }\href
  {https://www.cambridge.org/core/books/quantum-optics/1A5998427FD2BCA26B9C4FA026E84A1E}
  {\emph {\bibinfo {title} {Quantum Optics}}}\ (\bibinfo  {publisher}
  {Cambridge University Press},\ \bibinfo {year} {1997})\BibitemShut {NoStop}%
\bibitem [{\citenamefont {Vishwakarma}\ \emph {et~al.}(2023)\citenamefont
  {Vishwakarma}, \citenamefont {Sonpal}, \citenamefont {Pradhan}, \citenamefont
  {Haghighatlari}, \citenamefont {Afzal},\ and\ \citenamefont
  {Hachmann}}]{VISHWAKARMA2023653}%
  \BibitemOpen
  \bibfield  {author} {\bibinfo {author} {\bibfnamefont {G.}~\bibnamefont
  {Vishwakarma}}, \bibinfo {author} {\bibfnamefont {A.}~\bibnamefont {Sonpal}},
  \bibinfo {author} {\bibfnamefont {A.}~\bibnamefont {Pradhan}}, \bibinfo
  {author} {\bibfnamefont {M.}~\bibnamefont {Haghighatlari}}, \bibinfo {author}
  {\bibfnamefont {M.~A.~F.}\ \bibnamefont {Afzal}},\ and\ \bibinfo {author}
  {\bibfnamefont {J.}~\bibnamefont {Hachmann}},\ }in\ \href
  {https://doi.org/https://doi.org/10.1016/B978-0-323-90049-2.00028-7} {\emph
  {\bibinfo {booktitle} {Quantum Chemistry in the Age of Machine Learning}}},\
  \bibinfo {editor} {edited by\ \bibinfo {editor} {\bibfnamefont {P.~O.}\
  \bibnamefont {Dral}}}\ (\bibinfo  {publisher} {Elsevier},\ \bibinfo {year}
  {2023})\ pp.\ \bibinfo {pages} {653--674}\BibitemShut {NoStop}%
\end{thebibliography}
%

\end{document}